\documentclass[12pt,preprint]{aastex}
\usepackage{emulateapj5}
\usepackage{apjfonts}
\usepackage{epsfig}
\usepackage{natbib}
\newcommand{\alp}{\ensuremath{\alpha}}

\newcommand{\feh}{[Fe/H]}
\newcommand{\feave}{\ensuremath{\langle {\rm Fe} \rangle}}

\def\gtrsim{\mathrel{\hbox{\rlap{\hbox{\lower4pt\hbox{$\sim$}}}\hbox{\raise2pt\hbox{$>$}}}}}

\newcommand{\hbeta}{H\ensuremath{\beta}}

\newcommand{\kms}{km~s\ensuremath{^{-1}}}

\newcommand{\mgb}{Mg$b$}
\newcommand{\mgfe}{[Mg/Fe]}

\newcommand{\msun}{\ensuremath{M_{\odot}}}

\newcommand{\oii}{[\ion{O}{2}]}
\newcommand{\oiii}{[\ion{O}{3}]}

\newcommand{\sigmastar}{\ensuremath{\sigma_{\ast}}}

\def\lax{{$\mathrel{\hbox{\rlap{\hbox{\lower4pt\hbox{$\sim$}}}\hbox{$<$}}}$}}
\def\gax{{$\mathrel{\hbox{\rlap{\hbox{\lower4pt\hbox{$\sim$}}}\hbox{$>$}}}$}}

\slugcomment{Feb 19, 2012;
accepted by {\it The Astrophysical Journal}.}
\shorttitle{{\it Elliptical Outskirts}}
\shortauthors{GREENE, ET AL.}

\begin{document}

\title{The Stellar Halos of Massive Elliptical Galaxies}

\author{Jenny E. Greene\altaffilmark{1,2}, Jeremy D. Murphy\altaffilmark{2},
Julia M. Comerford\altaffilmark{2}, Karl Gebhardt\altaffilmark{2},
 Joshua J. Adams\altaffilmark{2,3}}

\altaffiltext{1}{Department of Astrophysics, Princeton University, }
\altaffiltext{2}{Department of Astronomy, UT Austin, 1 University Station C1400, 
Austin, TX 71712}
\altaffiltext{3}{Observatories of the Carnegie Institution of Washington, 813 Santa
Barbara Street, Pasadena, CA 91101}

\begin{abstract}
  We use the Mitchell Spectrograph (formerly VIRUS-P) on the McDonald
  Observatory 2.7m Harlan J. Smith Telescope to search for the
  chemical signatures of massive elliptical galaxy assembly. The
  Mitchell Spectrograph is an integral-field spectrograph with a
  uniquely wide field of view ($107 \times 107$ sq arcsec), allowing
  us to achieve remarkably high signal-to-noise ratios of $\sim 20-70$
  per pixel in radial bins of $2-2.5$ times the effective radii of the
  eight galaxies in our sample. Focusing on a sample of massive
  elliptical galaxies with stellar velocity dispersions
  \sigmastar$>150$~\kms, we study the radial dependence in the
  equivalent widths (EW) of key metal absorption lines.  By twice the
  effective radius, the \mgb\ EWs have dropped by $\sim 50\%$, and only
  a weak correlation between \sigmastar\ and \mgb\ EW remains.  The \mgb\
  EWs at large radii are comparable to those seen in the centers of
  elliptical galaxies that are $\sim$ an order of magnitude less
  massive.  We find that the well-known metallicity gradients often observed 
  within an effective radius continue smoothly to $2.5 R_e$, while the
  abundance ratio gradients remain flat. Much
  like the halo of the Milky Way, the stellar halos of
  our galaxies have low metallicities and high \alp-abundance ratios,
  as expected for very old stars formed in small stellar systems.  Our
  observations support a picture in which the outer parts of
  massive elliptical galaxies are built by the accretion of much smaller
  systems whose star formation history was truncated at early times.
\end{abstract}

\section{Introduction}

Elliptical galaxies are comprised of mostly old stars, contain little
gas or dust, and show very tight scaling relations between their
sizes, central surface brightnesses, and stellar velocity dispersions
\citep[the Fundamental Plane,
e.g.,][]{djorgovskidavis1987,dressleretal1987}. Despite their apparent
simplicity, observations of elliptical galaxies continue to surprise
and confound us.  Their central stellar populations
suggest that the most massive elliptical galaxies formed their stars
rapidly and early \citep[e.g.,][]{faber1973,thomasetal2005}.  And yet,
evidence for dramatic size evolution has emerged, such that elliptical
galaxies at $z \approx 1$ were apparently a factor of $\sim 2$
smaller at fixed mass than they are today
\citep[e.g.,][]{trujilloetal2006,vandokkumetal2008,
  vanderweletal2008,cimattietal2008,damjanovetal2009,cassataetal2010}.
It is, of course, extremely challenging to measure galaxy sizes at
high redshift \citep[e.g.,][]{hopkinsetal2009,saraccoetal2010}, but
evidence continues to mount that the size evolution is real
\citep[e.g.,][]{newmanetal2011,brodieetal2011,papovichetal2011}.

The most common scenario to explain the dramatic size growth in
elliptical galaxies at late times invokes minor merging that can make
galaxies fluffier without adding very much mass 
\citep[e.g.,][]{gallagherostriker1972,boylankolchinma2007,naabetal2007,naabetal2009,
newmanetal2011}. Naively, late-time merging with small systems would
wash out the well-established scaling relations between stellar
velocity dispersion (\sigmastar) and stellar population properties
observed in local elliptical galaxies, such as the \mgb-\sigmastar\ relation
\citep[e.g.,][]{benderetal1993}.  Furthermore, stellar
population studies of local elliptical galaxies clearly find that the
stars in the most massive elliptical galaxies were formed earliest
($z>2$) and fastest ($<$~Gyr), while lower-mass systems have more
extended formation histories and later formation times
\citep[e.g.,][]{thomasetal2005}.  From the tight color-magnitude
relation alone it is hard to support much late-time star formation 
\citep[or the addition of more metal-poor stars, e.g.,][]{boweretal1992}.

The tension between the tight scaling relations of elliptical galaxies
and their apparent puffing up from late-time merging is alleviated if
the stars added at late times are deposited at large radius. The vast
majority of stellar population work is heavily weighted towards the
very luminous central component of these galaxies, usually well within
the half-light radius \citep[$R_e$;
e.g.,][]{faber1973,rawleetal2008,gravesetal2009,kuntschneretal2010}.  
Recently, thanks to the Sloan
Digital Sky Survey \citep[SDSS;][]{yorketal2000}, very large samples
of elliptical galaxies are now available for examining color gradients
\citep[e.g.,][]{zibettietal2005,tortoraetal2010,suhetal2010,gonzalez-perezetal2011}.
With few exceptions \citep[e.g.,][]{rudicketal2010}, these
observations have not extended much beyond the effective radius.  To
fully exploit the fossil record to understand the assembly of
elliptical galaxies, we ought to look for radial changes in the
stellar population, particularly beyond $R_e$.

The study of the radial dependence of chemical composition in
elliptical galaxies has a long history.  Imaging studies of elliptical
galaxy colors date back to \citet{devaucouleurs1961}.  Since then,
there have been many studies made of the radial color gradients in
elliptical galaxies \citep[e.g.,][]{tifft1969,wirthshaw1983,eisenhardtetal2007}.
The summary presented in \citet{stromstrom1978} remains accurate today;
elliptical galaxies are bluer at large radii, most likely due to a
decline in metallicity \citep{spinrad1972,strometal1976}. 
However, with photometry alone it is difficult to precisely
disentangle the well-known degeneracies between age and metallicity
\citep[e.g.,][]{Worthey1994}.  Many spectroscopic surveys have looked
at the gradients in the equivalent widths (EW) of key metal lines that
can break these degeneracies
\citep[e.g.,][]{spinradtaylor1971,faberetal1977,gorgasetal1990,
  fisheretal1995,mehlertetal2003,ogandoetal2005,broughetal2007,
  baesetal2007,annibalietal2007,sanchez-blazquezetal2007,rawleetal2008,
  kuntschneretal2010,weijmansetal2009}. Roughly speaking,
spectroscopic work confirms the overall conclusions from 
  imaging studies. Metallicity dominates the color changes,
decreasing outwards by 0.1-0.5 dex per decade in radius. In general
there is no strong evidence for age gradients (although see Baes et
al. for an alternate view). 

\begin{figure*}
\vbox{ 
\vskip -5mm
\hskip +0mm
\psfig{file=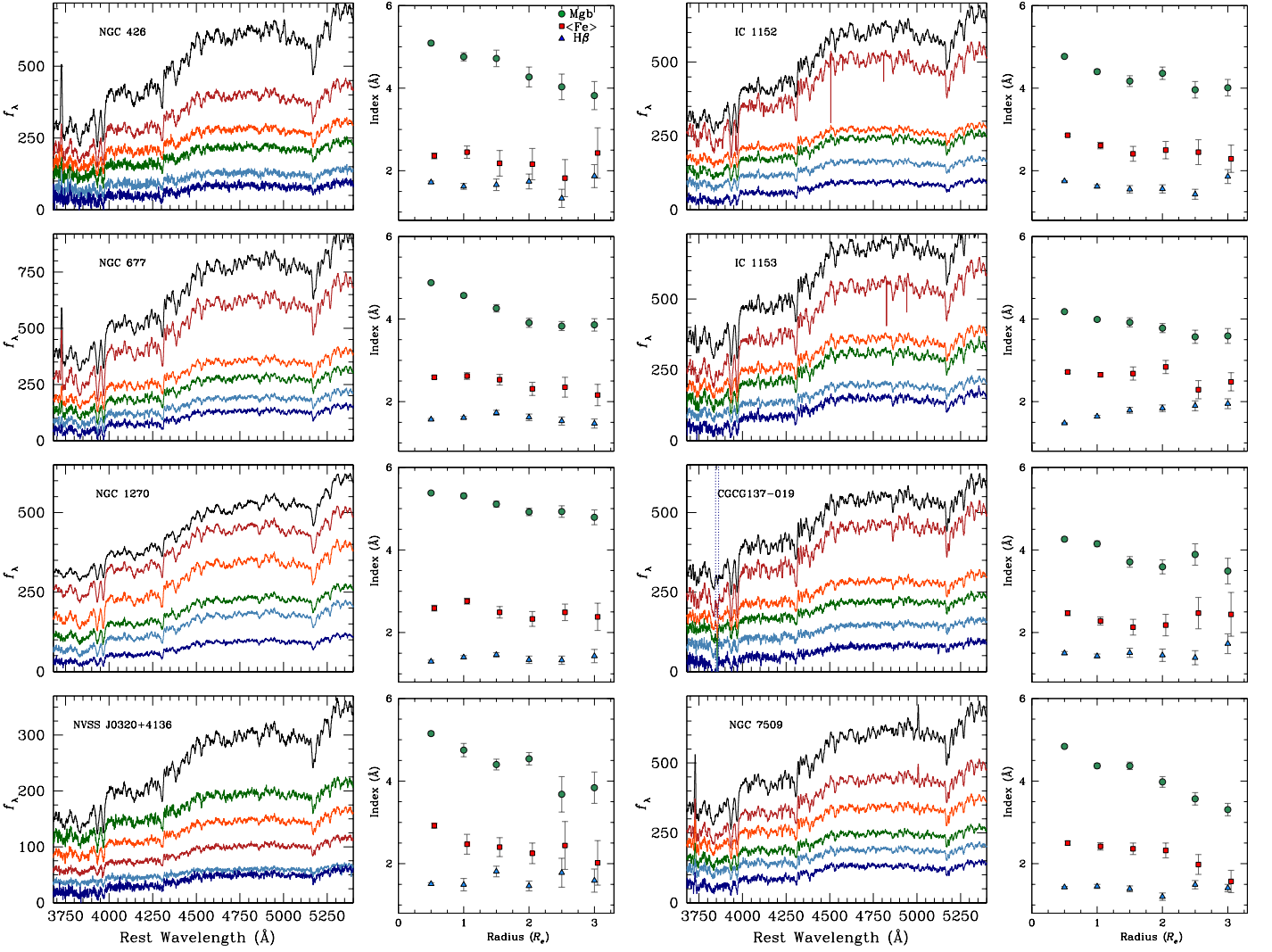,width=1.0\textwidth,keepaspectratio=true,angle=0}
}
\vskip -0mm
\figcaption[]{
 Spectra (left, middle) and index equivalent widths (right) for our 
  sample.  {\it Left, middle}: Spectra are
  plotted in elliptical annuli of (from top to bottom) $0-0.5 R_e$, $0.5-1 R_e$, 
  $1-1.5 R_e$, $1.5-2 R_e$, and $2-2.5 R_e$ (see Table 1). Units are
  $10^{-17}$erg~s$^{-1}$~cm$^{-2}$~\AA$^{-1}$, but the spectra have
  been offset for clarity. The sky feature in CGCG137$-019$ is highlighted with a 
  dotted vertical line.
  {\it Right}: We show the \mgb\ index (green
  circles), the $\langle$Fe$\rangle$ index, which is the
average of the Fe 5270 and Fe 5335 indices (red
  squares), and the H$\beta$ index (blue triangles) in \AA. Error bars
  are derived via Monte Carlo simulations as described in \S 4.2.1.
  The $\langle$Fe$\rangle$ index is offset slightly in radius for
  clarity.
\label{fig:spec_pg1}}
\end{figure*}

Spectroscopic surveys can track more than just metallicity and age.
They can also study the relative abundances of individual elements. In
particular, the \alp\ elements (e.g., Mg, C, O, N) are formed in Type
II supernova explosions, while the Fe-peak elements (Fe, Cr, Mn) are
formed predominantly in Type Ia supernovae, and are thus produced with
a temporal lag from the peak of star formation.  The relative
quantity of \alp\ to Fe-peak elements provides a star-formation timescale,
with enhanced \alp/Fe ratios pointing to rapid time-scales of star
formation.  Elliptical galaxies display a strong trend of increasing
\alp/Fe abundance with increasing mass or stellar velocity dispersion
\citep[e.g.,][]{faber1973,terlevichetal1981,wortheyetal1992} although
see also \citet{kelsonetal2006}.  It is therefore thought that the most massive 
elliptical galaxies formed their stars rapidly and at $z \gtrsim 2$ 
\citep[e.g.,][]{thomasetal2005}. Thus far, no strong
gradients in \alp/Fe ratios have been detected at large radii
\citep[e.g.,][and references therein]{kuntschneretal2010,spolaoretal2010}.

A few studies have managed to probe stellar populations in elliptical
galaxies beyond the effective radius.  It is very hard to achieve the
required signal-to-noise at large radii with long-slit spectroscopy
since the area subtended on the sky is small and the sky level is
factors of several brighter than the signal
\citep[e.g.,][]{kelsonetal2002,sanchez-blazquezetal2007}.
Integral-field unit (IFU) spectroscopy provides two-dimensional
information, and coadding the signal in annuli strongly boosts the
signal relative to the sky.  A handful of studies thus far have used
IFUs with smaller fields of view, and either tile the instrument at
large radius \citep[e.g.,][]{weijmansetal2009} or focus on the central
regions of the galaxy \citep[][]{rawleetal2008,kuntschneretal2010}. In
this work, we exploit the $4\farcs2$ diameter fibers and
$107\times107$\arcsec\ field of view of the Mitchell Spectrograph to
study the spatial variation in age, metallicity, and abundance ratio
gradients for eight massive early-type galaxies.  Our increased leverage on 
stellar populations in the galaxy outskirts will allow us to put new constraints 
on the assembly of massive elliptical galaxies at late times.

In \S 2 we describe the sample and in \S 3 we describe the instrument and
data reduction.  The analysis is described in \S 4.  Those most interested in 
results can focus on \S 5 and the subsequent discussion in \S 6.
We summarize and conclude in \S 7.  When needed, we use the
standard concordance cosmology of \citet[][]{dunkleyetal2009}.

\section{Sample}
\label{sec:Sample}

We start with a small pilot sample of eight galaxies as a proof of
concept that the Mitchell Spectrograph is well-suited to this work
(Table 1, Figure \ref{fig:spec_pg1}).  The sample selection is not
ideal, and we do not make claims of its completeness or uniformity,
since the galaxies were selected with other science goals in mind.  In
short, we selected galaxies with red colors
\citep[$u-r>2.2$;][]{stratevaetal2001}, stellar velocity dispersions 
that are larger than the instrumental resolution
of the Mitchell Spectrograph (\sigmastar$> 150$~\kms; as measured by
the SDSS pipeline) and redshifts in the narrow range 
\vbox{ 
\hskip 0.in
\psfig{file=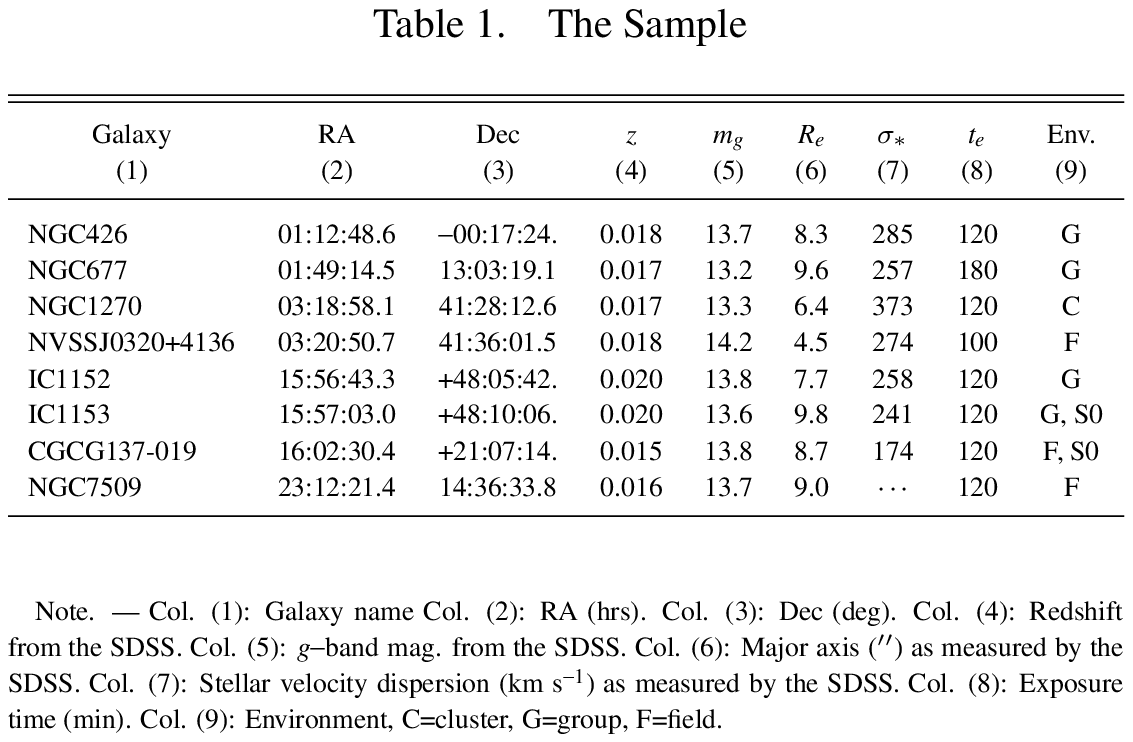,width=0.45\textwidth,keepaspectratio=true,angle=0}
}
\vskip 4mm
\noindent
$0.015 < z < 0.02$ (85 Mpc, for a scale of 0.4 kpc per \arcsec).  We examine all of
the candidates and remove obvious edge-on disk galaxies, but have made
no formal morphology cut.  Thus there are two S0s in the final sample
(CGCG137$-$019 and IC 1153).  We also make no selection on
environment.  However, in the sample there is a cluster galaxy
\citep[NGC 1270; ][]{millerowen2001}, two brightest group galaxies
\citep[NGC 677 and NGC 426;][]{berlindetal2006}, two that belong
to the same group \citep[IC 1152 and IC 1153;][]{whiteetal1999} and the
rest are in lower density environments.  Using our stellar velocity dispersion 
measurements, simple dynamical mass estimates for the galaxies 
range from $8 \times 10^{10}-3 \times 10^{11}$~\msun, with only one 
below $10^{11}$~\msun.  Based on the mass function from \citet{belletal2003}, 
they range from one to four times $M^*$ for ellipticals.

\section{Observations and Data Reduction}
\label{sec:obs}

The observations were obtained over two runs, one in Sept 2010
(including the bulk of the galaxies) and the other in June 2011 (IC
1152, IC 1153; Table 1).  We used the George and Cynthia Mitchell
Spectrograph \citep[the Mitchell Spectrograph, formerly
VIRUS-P;][]{hilletal2008a} on the 2.7m Harlan J. Smith telescope at
McDonald Observatory. The Mitchell Spectrograph was built as a
prototype for the VIRUS spectrograph that will soon be deployed on the
Hobby-Eberly Telescope to perform a dark energy experiment
\citep[HETDEX;][]{hilletal2008b}.  Each of the 246 fibers subtends
$4\farcs2$ and are assembled in an array similar to Densepak
\citep{bardenetal1998} with a 107\arcsec$\times$107\arcsec\ field of
view and a one-third filling factor. The Mitchell Spectrograph has
performed a very successful search for Ly$\alpha$ emitters
\citep{adamsetal2011,finkelsteinetal2011,blancetal2011} and has become
a highly productive tool to study spatially resolved kinematics and
stellar populations in nearby galaxies
\citep{blancetal2009,yoachimetal2010, murphyetal2011,adamsetal2012}.

We used the low-resolution blue setting of the Mitchell
Spectrograph. Our wavelength range spans 3550-5850\AA\ with an
average spectral resolution of $5\,$\AA\ FWHM. This resolution
delivers a dispersion of $\sim$~$1.1\,$\AA\ pixel$^{-1}$ and corresponds
to \sigmastar~$\approx 150$~\kms\ at $4300\,$\AA, our bluest Lick index.  Each
galaxy was observed for a total of $\sim 2$ hours with one-third of
the time spent at each of three dither positions to fill the field of
view.  Initial data reduction was accomplished using the custom code
Vaccine \citep{adamsetal2011,murphyetal2011}.  We briefly review the
steps of the pipeline here, but refer the interested reader to the
previous papers for more detailed discussion. Initial overscan and
bias subtraction are performed first on all science and calibration
frames. All co-additions of data and calibration frames are performed
with the biweight estimator (Beers et al. 1990). Twilight flats are
used to construct a trace for each fiber, which takes into account
curvature in the spatial direction. We employ a routine similar to
that proposed by \citet{kelson2003} to avoid interpolation during this
step. Thanks to this special care, correlated errors are avoided and
it is possible to track the S/N in each pixel through the remainder of
the reductions. Knowing the S/N in each pixel enables deeper limits in detection
experiments such as those performed by \citet{adamsetal2011}. All
subsequent operations are conducted in the new trace coordinate system
within a cross-dispersion aperture of 5 pixels.

To correct for curvature in the spectral direction, a wavelength solution
is derived for each fiber based on arcs taken both at the start and
end of the night. Gaussian fits to known arc lines are fit with a
fourth-order polynomial to derive a complete wavelength solution for
each fiber. The typical rms residual variations about this
best-fit fourth-order polynomial are $0.08\,$\AA\ for the Sept 2010
data and $0.04\,$\AA\ for our June 2011 data. A heliocentric correction
is then calculated for each science frame.

\begin{figure*}
\vbox{ 
\vskip -3mm
\hskip +10mm
\psfig{file=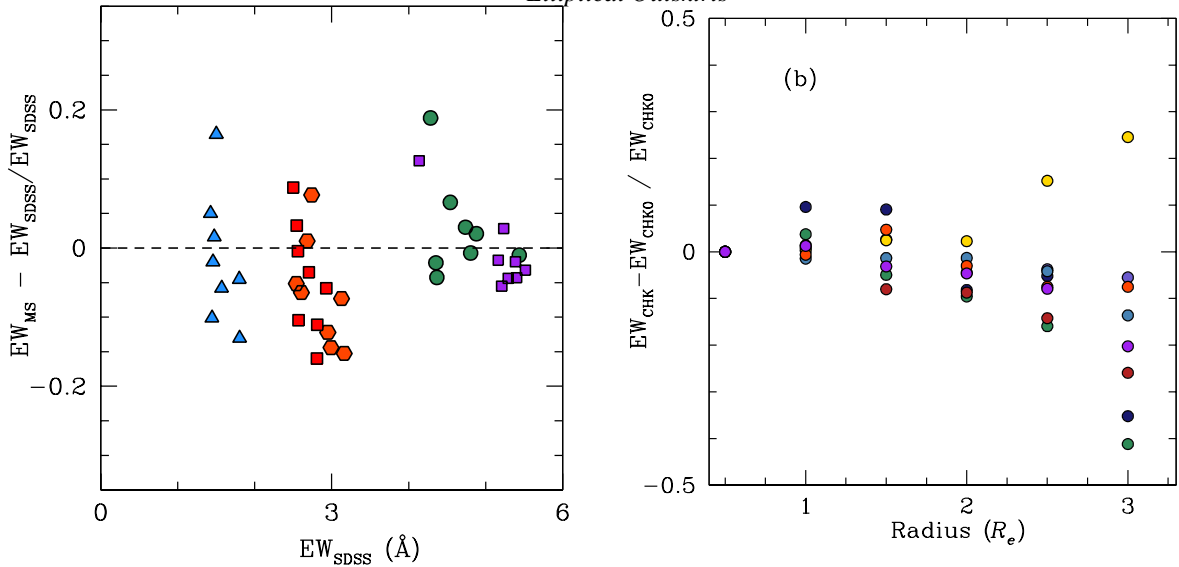,width=0.4\textwidth,keepaspectratio=true,angle=-90}
}
\vskip -0mm
\figcaption[]{
{\bf a}: 
Comparison between the Lick indices measured from the SDSS spectrum
and our central 0.5$R_e$ bin for \hbeta\ (blue
triangles), the G-band at $4300 \,$\AA\ (purple squares), the \mgb\ triplet
(green circles), the Fe 5270 line (red squares), and the Fe 5335 line 
(red hexagons).  Fractionally, the
two sets of indices agree within $3\%$ of each other, while the mean
deviation $\langle ({\rm EW}_{\rm MS} - {\rm EW}_{\rm SDSS})/{\rm EW}_{\rm
  SDSS} \rangle = -0.02 \pm 0.08$.  The $\langle$Fe$\rangle$ 
index is systematically offset to lower values in our spectra 
by enough to introduce significant systematic offsets in
the inferred metallicity.
{\bf b}: Fractional change in the EW of Ca H+K as a function of radius
for all galaxies in the sample.  Each color represents a different
source.  Note that the fluctuations between $0<$R$_e<2$ are within
$10\%$, which is both the typical error in these measurements and the
level at which Ca H+K is expected to vary due to real changes in
stellar population.  It is only in the outermost bin (2.5-3$R_e$) that
we see systematic effects begin to artificially lower the EW in some
objects.  The yellow points that rise at large radii correspond to
CGCG137$-019$, which was observed at twilight and has imperfect sky
subtraction.
\label{fig:cflick}}
\end{figure*}

Next, a flat field is constructed from the twilight flats taken at
both the start and end of the night. Variations in temperature never
exceeded 2~C for any of our observing nights and the stability of the
flat field has been shown to be $<0.1$ pixels under these conditions
\citep{adamsetal2011}. As the twilight flats contain solar spectrum,
we generate a model of this component by employing a bspline fitting
routine \citep{dierckx1993}. A boxcar of 51 fibers is employed to
model the solar spectrum that effectively removes all cosmic rays,
continuum sources, and variations in the flat field, in order to
isolate the solar spectrum. The spatial and spectral curvature are
leveraged here in order to supersample the solar spectra within the
boxcar. This supersampled bspline fit to the solar spectra within each
fiber is then divided into the original flat. What remains are the
flat field effects that we want to capture: variations in the
individual pixel response, in the relative fiber-to-fiber variation, and
in the cross-dispersion profile shape for every fiber. This flat field
is then applied to all of the science frames.

The next step is sky subtraction. Unlike some instruments \citep[e.g.,
Sauron;][]{baconetal2001}, the Mitchell Spectrograph does not have
dedicated sky fibers. Instead, we observed off-galaxy sky frames with
a sky-object-object-sky pattern, with five minute exposure times on
sky and twenty minute object exposures. The sky nods are processed in
the same manner as the science frames described above. In order to
create a sky frame we give an equal weighting of two to each sky nod,
then coadd them to achieve an equivalent exposure time as the science
frames. The advantage of sky nods is the high S/N we achieve in our
sky estimate, based on the large number of fibers in the IFU. The
disadvantage is that we sample the sky at a different time than the
science frames and are thus subject to temporal variations in the
night sky, particularly at twilight. However, by varying the weights
given to each sky nod we are able to explore possible systematics due
to sky variability. For all lines measured, no EW value changes by
more than $0.08\,$\AA\ due to temporal sky changes, 
even at the largest radii considered in this 
paper. We give the details and return to possible
systematics related to sky subtraction in \S 4.2.1. Once the sky
subtraction is complete, cosmic rays are identified and masked.

We use software developed for the VENGA project \citep{blancetal2009}
for flux calibration and final processing.  We observe flux
calibration stars each night using a six-point dither pattern and
derive a relative flux calibration in the standard way.  Then we use
tools developed by M.~Song, et al. (in preparation) to derive an
absolute flux calibration relative to the SDSS imaging.  M. Song uses
synthetic photometry on each fiber and scales it to match the SDSS
$g-$band image of each field, with a median final correction of $\sim
20\%$.  The correction exceeds $50\%$ only during a period of high
cirrus in the second night of observing in Sept 2010, which affects
NVSS J0320+4136 and NGC 426.  Finally, all fibers are interpolated
onto the same wavelength scale and combined.  

\subsection{Radial bins}

We focus on spectra combined in elliptical annuli with a width of
0.5$R_e$.  Since the effective radii of these galaxies are $\sim$
twice the fiber diameter of 4\arcsec, 0.5$R_e$ is roughly the scale 
of a single fiber.  In all cases we use the de Vaucouleurs radius derived by the
SDSS pipeline.  Since we are averaging over such large physical areas
on the sky, the exact measured $R_e$ should not impact the
conclusions. We did experiment with using wider radial bins at large
radius to extend further from the galaxy center.  However, we did not
boost the S/N appreciably in this way.  Furthermore, we begin to be
systematics limited at $\sim 3 R_e$ (\S 4.2.1).  The bins consist of
1-4 fibers at $0-0.5 R_e$, increasing to 20-40 fibers at $2-2.5
R_e$. The S/N per final coadded spectrum is shown in Table 2.

Since all of our galaxies by selection have SDSS spectra, we can test
the wavelength dependence of the flux calibration by comparing the 
shape of the spectrum in the central fiber of the Mitchell
Spectrograph with the SDSS spectrum.  We find $\sim 5\%$ agreement in
nearly all cases, with no more than $\sim 15\%$ differences at worst.

\section{Analysis}
\label{sec:Analysis}

Our ultimate goal is to derive the stellar population properties of the 
galaxies using the absorption line spectra. In principle, we can use
full spectral synthesis techniques
\citep[e.g.,][]{bruzualcharlot2003,coelhoetal2007,vazdekisetal2010},
and exploit all the information available in the spectra.
However, in practice, there are a number of hurdles, including
imperfect flux calibration and systematic color effects, that make the
model fitting sensitive in systematic ways to imperfections in our
data. It is known that the models are not always able to fit the
absorption line equivalent widths \citep[e.g.,][]{gallazzietal2005},  
although see also \citet{kolevaetal2011}.
Since we are most interested in measuring changes in stellar
population properties as a function of radius, we use line index
measurements \citep[e.g., Lick
indices,][]{faberetal1985,wortheyetal1994}. As most prior work on
this subject has used similar methodology, we are in a good position
to compare with the literature.  Keep in mind that we are thus measuring the
luminosity-weighted mean properties of the stellar population.

Another major uncertainty in these stellar population synthesis models
comes from our relative ignorance of stellar spectral energy
distributions for stars with very different metallicities and/or
abundance ratio patterns than stars in our solar neighborhood.  These
modeling deficiencies impact our study directly, since elliptical
galaxies tend to have higher metallicities and \alp-abundance ratios than
local stars.  Lately, the problem has garnered substantial attention
both from the point of view of full spectra synthesis
\citep[e.g.,][]{coelhoetal2007,conroyvandokkum2011,
  marastonstromback2011} and for Lick index inversion methods
\citep{thomasetal2003,schiavon2007,vazdekisetal2010}.  We exploit
these modern models, but note that our knowledge of the underlying
stellar evolution remains imperfect.

\subsection{Emission line contamination}

Something like $80\%$ of elliptical galaxies contain low levels of
ionized gas within an effective radius
\citep{sarzietal2010,yanblanton2011}.  For our purposes, this emission
serves only as a contaminant, as it fills in the absorption features
that we are trying to measure.  By far the strongest emission feature
in our spectra is the \oii$~\lambda \lambda 3726, 3729$ line, but 
there are no absorption features of interest that are confused by \oii.
It is contamination from \hbeta\ and \oiii$~\lambda 5007$ 
that concerns us here.
In order to correct for this low-level emission, we adapt the code
pPXF+GANDALF developed by M. Sarzi \citep{sarzietal2006} and
M. Cappellari \citep{cappellariemsellem2004}.  
pPXF performs a
weighted fit to the galaxy continuum using spectral templates provided
by the user, including a polynomial fit to the continuum and a Gaussian
broadening to represent the intrinsic dispersion of the
galaxy. From these fits we derive a measurement of the stellar velocity dispersion 
in each radial bin. GANDALF iteratively measures the emission and continuum
features simultaneously, to achieve an unbiased measurement of both
components.  As templates, we use \citet{bruzualcharlot2003}
single-age stellar population models with $\sigma \approx 70$~\kms\
resolution.  In principle, we could use these fits to trace the
stellar populations with radius, but for the reasons discussed above,
we do not find this methodology robust.

In general, the emission-line EW is small compared to that of the
absorption lines.  The maximum contamination comes in the central
fiber of NGC 7509 (which has high-ionization lines indicative of an
accreting black hole at the center).  However, note that our spectral
resolution of $150$~\kms\ is not necessarily high enough to resolve
the emission lines. 

\vbox{ 
\hskip 0.in
\psfig{file=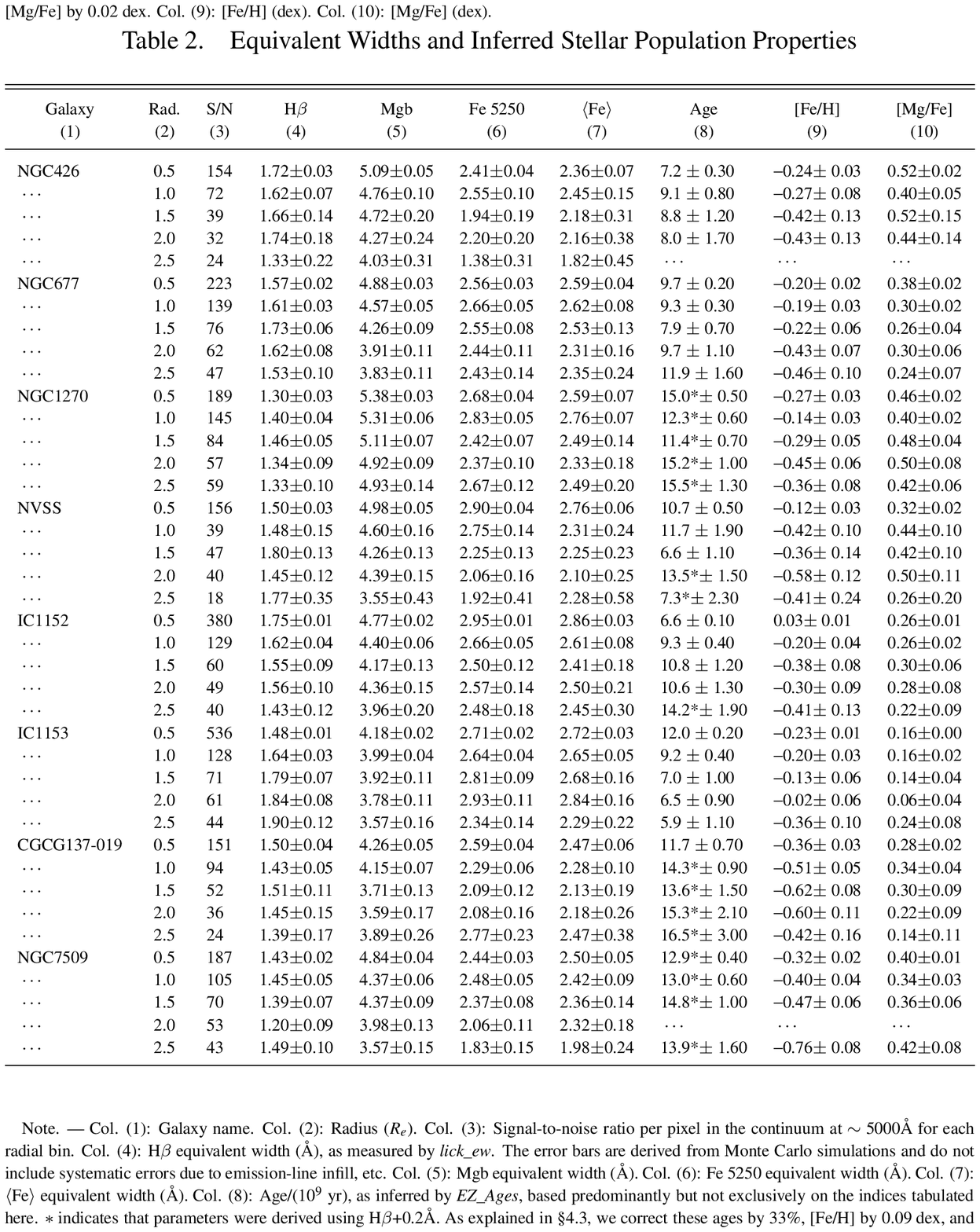,width=0.45\textwidth,keepaspectratio=true,angle=0}
}
\vskip 4mm
\noindent
We compare the fit to \oiii\ from the central pixel of the 
Mitchell spectra with the SDSS line fits at twice the resolution.  
The two fits agree within $\sim 30\%$ (range of $2-70\%$ 
differences).  As a result of the weakness of the features
and the low spectral resolution, we do not have strong constraints on
the flux or line shape of these emission lines. We
incorporate emission line subtraction into our error bar estimates as
described below.  While the emission-line gas is not the focus of this
paper, we note that in a few cases the \oii\ emission is observed to
extend beyond $2 R_e$.  An analysis of this emission will be the focus
of a different work.

Unfortunately, even very small corrections do lead to substantive
changes in the inferred galaxy ages -- $0.1\,$\AA\ of \hbeta\ emission
infill corresponds to an age difference of $\sim 1$ Gyr
\citep{schiavon2007,gravesfaber2010}.  Thus, we caution that the
absolute ages derived here are uncertain at this level.  In some
cases, in fact, the small infill of the \hbeta\ lines by emission
leads to \hbeta\ EWs that are smaller than even the oldest and most
metal-rich single stellar population (SSP) models.
\citet{gravesfaber2010} describes a detailed method to detect \oiii\
at the $0.2\,$\AA\ level.  As described below, we do not have the S/N to
perform their analysis, but the level of infill will become relevant
again when we derive stellar populations in \S 4.3.1.  

\subsection{Equivalent Widths}

Measuring absorption-line indices and placing them on a common system
is a delicate art that is sensitive not only to the spectral
resolution of the instrument, but also flux calibration and S/N
\citep[e.g.,][]{wortheyetal1994,schiavon2007,yan2011}.  We utilize the
flexible and robust IDL code {\it lick\_ew}, written by G. Graves
\citep{gravesschiavon2008}.  The code takes as input the stellar velocity 
dispersion at each radial bin and puts the measured EWs on the Lick 
system. In principle we measure EWs for all 26
Lick indices, but we focus our attention on \hbeta, \mgb, Fe 5270, and
Fe 5335 (Table 2).  We also measure the Ca H+K index as defined by
\citet{brodiehanes1986}.  Because of its high EW, any change in Ca H+K
EW are indicative of systematic effects in the spectra (see \S
  4.2.1).

Since we compare with Lick indices defined on flux-calibrated stars
\citep{schiavon2007}, we only correct the indices to a standard spectral 
resolution but apply no other zeropoint offsets. This same approach is taken by
\citet{gravesetal2009}.  By comparing the Mitchell indices from the
central spectrum with the indices measured from the SDSS spectra, we
confirm that we are on the same system (Figure \ref{fig:cflick}).
Note that the apertures are not perfectly matched (3\arcsec\ vs
4\farcs2) but the difference between these two should be small since
the observed gradients are gentle.  Spectra and Lick indices for all
galaxies are presented in Figure \ref{fig:spec_pg1} and Table 2.  We
find decent agreement between our indices and those from the SDSS
spectra, except in the case of the Fe indices.  Taking average
differences $\Delta {\rm Index} = \langle {[\rm Index_{Mitchell} -
  Index_{SDSS}]/Index_{SDSS}}\rangle$, and the standard deviation
therein, we find $\Delta$Mg{\it b}$=0.03 \pm 0.07$, $\Delta$\hbeta$=-0.02
\pm 0.09$, $\Delta$Fe 5270$=-0.07 \pm 0.08$, $\Delta$Fe 5335$=-0.04
\pm 0.08$, and an overall offset of $\Delta$ Index$=-0.02 \pm 0.08$
that also includes the G-band.

If instead we look at absolute differences, we find that there is a
small systematic difference in the Fe indices, in the sense that
$\langle$Fe$_{\rm MS}\rangle- \langle$Fe$_{\rm SDSS}\rangle =
-0.14\,$\AA.  While the systematic offset in $\langle$Fe$\rangle$ is
small, it translates to large systematic errors of $0.08$ dex in \feh\
and $0.06$ dex in \mgfe.  We have explored various causes for the 
systematic offset, including non-Gaussian line-broadening functions,
variations in resolution with wavelength, and sky subtraction.  None alone 
is sufficient to explain this systematic effect, although likely a combination 
of these, and possibly small-scale errors in flux-calibration, are to blame.
As described above, our \hbeta\ measurements are
likely suffering from very low levels of emission-line infill, which
makes it difficult to derive absolute ages.  Thus, we do not focus on
the absolute values of the derived parameters here, but rather on our
main strength, which is gradients out to large radii.

\begin{figure*}
\vbox{ 
\vskip 20mm
\hskip +0.5in
\psfig{file=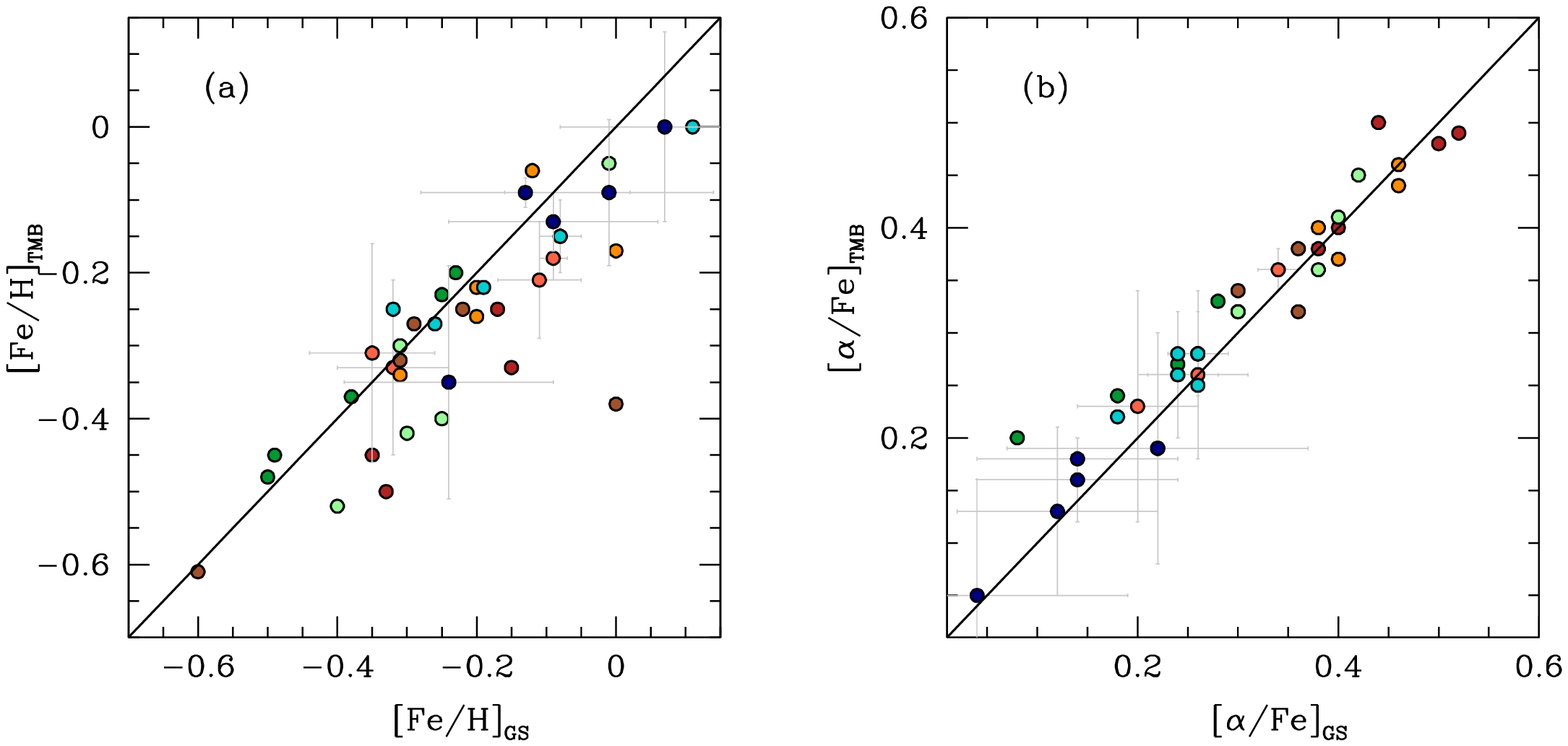,width=0.4\textwidth,keepaspectratio=true,angle=-90}
}
\vskip -0mm
\figcaption[]{
We compare metallicities (\feh; {\bf a}) and $\alpha$ abundances ([\alp/Fe]
as measured by the Mgb index; {\bf b}) derived 
using the prescriptions of \citet{gravesschiavon2008} and \citet{thomasetal2003} 
respectively.  We show observations for all galaxies at radii out to $2.5 R_e$.  
Each galaxy is represented by a different color.  The agreement between 
the two models is decent, with 
$\langle$\feh$_{\rm GS} -$\feh$_{\rm TMB} \rangle = -0.06 \pm 0.1$, 
and $\langle$[\alp/Fe]$_{\rm GS } -$[\alp/Fe]$_{\rm TMB} \rangle = -0.03 \pm 0.08$.  
To reduce crowding, we include error bars for only two galaxies in the sample.
\label{fig:cftmbgraves}}
\end{figure*}
\vskip 5mm

\subsubsection{Uncertainties}

Potential contributors to our error budget include random noise,
emission line removal, and sky subtraction.  We can build the former
two into Monte Carlo simulations of our measurement process.  For each
galaxy, in each radial bin, we start with the best-fit model from
GANDALF and create 100 realizations, using the error spectrum
generated by Vaccine.  We rerun GANDALF and {\it lick\_ew} on each
artificial spectrum and take the error bar as defined by the values
encompassing 68\% of the mock measurements.  Note that the model
contains both emission and absorption lines, and thus we should
include the uncertainties due to emission line removal naturally in
our error budget.

Sky subtraction is a potentially significant source of uncertainty
since we are working factors of a few below the level of the sky.  In
order to quantify how much sky variability effects our final science
results we take a heuristic approach.  We explore two different
scenarios. The first quantifies how changes in the night sky between
our two sky nods impacts our final measured EW values. The second test
quantifies the effect of an overall over- or under-subtraction of the
night sky. In both cases we allow the weighting given to each sky nod
to vary, then carry through all the resulting variations in the
subtracted science frames. A comparison of the final measured EW
values allows for a very direct measure of how well we are handling
sky subtraction. We give the details of both scenarios here.

To quantify how changes in the night sky between our two sky nods
influence our final measured EW values we explore a range of weighting
in the sky nods used for subtraction. For example, if the sky did not
vary over the $\sim 45$ minutes between our two sky nods, an equal
weighting of 2.0 given to each sky nod would be appropriate. This
equates to an equal amount of total exposure time for each sky nod
(e.g., 5 min $\times 2.0$ + 5 min $\times 2.0$ = 20 min). However, the
sky may evolve on timescales shorter than this. To quantify the effect
of an evolving sky on our final science results we ran several
different sets of sky nod weightings through our reduction routines
and compared the final EW values for all indices. We average
deviations over radial bins from $1.5-3 R_e$.  The sky weightings we explored varied by
$\pm 20\%$ from 2.0, yet always with a total weighting of 4.0
(e.g. 1.8 for sky nod 1 and 2.2 for sky nod 2). We then make direct
comparisons of the EW values ($\delta$EW = EW$_{\rm orig} -$EW$_{\rm
  new}$) for all of the lines. As the observing conditions were
predominantly stable for all our nights, we ran these tests on a
single galaxy (NGC~677) and believe it to be representative. The
largest $\delta$EW value measured was $0.080 \pm 0.098 \,$\AA\ (for
H$\beta$) when comparing a 2.2~-~1.8 to a 1.8~-~2.2 weighting. The
$\delta$EW for each line for the case above were as follows: Mgb =
$0.02 \pm 0.02 \,$\AA; H$\beta$ = $0.080 \pm 0.1 \,$\AA; Fe~5270 = $-0.007
\pm 0.02 \,$\AA. From an analysis of the variability of the sky spectra
over several nights \citep[Figure 15 in][]{murphyetal2011} the case of
20\% variability is extreme. The one exception occurred with IC~1152,
which saw a rising moon for some of the exposures. In this case we
explored a wide range of sky weights and found a weighting of 2.4~-~1.6
to be optimal.

The second scenario we explore is aimed at understanding what
systematic effect over- or under-subtraction of the sky has on our EW
values. To test this we conducted a similar set of tests, yet allowed
the final weighting of 4.0 to vary. We ran tests exploring both a 5\%
and 10\% over and under subtraction, relative to equal exposure
time. We then made the same comparison in $\delta$EW as described
above. In the case of a $\pm 5\%$ systematic error in subtraction, we
find $\delta$\mgb\ of $-0.09 \pm 0.04 \,$\AA, $\delta$\hbeta\ of $0.05 \pm
0.09 \,$\AA, and $\delta$<Fe> of $-0.05 \pm 0.02 \,$\AA.  The worst
deviations in individual bins are at the $0.15 \,$\AA\ level in \mgb\ and
\hbeta\ for the 5\% oversubtraction case.  It is interesting to note that there 
are not strong systematic effects.  Instead we see the indices bounce around 
at the $0.1-0.15 \,$\AA\ level for this level of sky subtraction error.
When we get to $10\%$ over-subtraction, the errors are
$\delta$\mgb\ of $-0.2 \pm 0.08 \,$\AA, $\delta$\hbeta\ of $0.01 \pm
0.04 \,$\AA, and $\delta$<Fe> of $-0.04 \pm 0.08 \,$\AA.  The worst
deviations in individual bins are at the $0.2 \,$\AA\ level in \mgb\ and
\hbeta\ for 10\% oversubtraction.  Larger fractional errors in sky level 
would lead to obvious residuals in our outer fibers that we do not see. 
As all of the $\delta$EW values calculated from both tests
described here are within our typical uncertainties, and the scenarios
we tested were extreme cases, we conclude that our results are robust
against sky variations on the scale of $\sim 45$ minutes seen in our
data set.

As an additional sanity check of our sky subtraction, we calculate the
EW of the Ca H+K $\lambda \lambda 3934, 3968$ lines.  These features
have very high EW, but also are virtually insensitive to changes in
stellar populations (at the $\sim 10\%$ level; \citet{brodiehanes1986}).
Thus, we expect the line depths to be constant with radius.  Because
these features are quite blue, they provide a rather stringent test of
our fidelity in the sense that the blue spectral shape is most
sensitive to errors in sky subtraction. We find that these lines do
not vary by more than $10\%$ out to $2.5 R_e$ for most systems (Figure
\ref{fig:cflick}{\it b}).  As a result, we view our results out to $2.5
R_e$ as reliable.  While we show the point at $2.5-3 R_e$ in the
figures, we will not use that point in our fitting.

\subsection{Stellar population modeling}

We now convert the observed EWs into ages, metallicities, and
abundance ratios using SSP models.  Since all
indices are blends of multiple elements and all depend on age,
metallicity, and abundance ratio to some degree
\citep[e.g.,][]{wortheyetal1994}, modeling is required to invert the
observed EWs and infer stellar population properties.  We compare two
different modeling techniques.  We first use the methodology outlined
in \citet{thomasetal2003,thomasetal2005}.  Since the EWs of \mgb\ and
\feave\ change with both [Fe/H] and [Mg/Fe], these authors construct
linear combinations of the two indices.  The index
[MgFe']$=\sqrt(\rm{Mg{\it b} [0.72 \, Fe 5270 + 0.28 \, Fe 5335]})$ is
independent of [Mg/Fe] and tracks [Fe/H], while the index Mg{\it b}/\feave\
depends only on [Mg/Fe]. The pair of derived indices is then inverted
to infer metal content and abundance ratios.

\begin{figure*}
\vbox{ 
\vskip 0mm
\hskip 20mm
\psfig{file=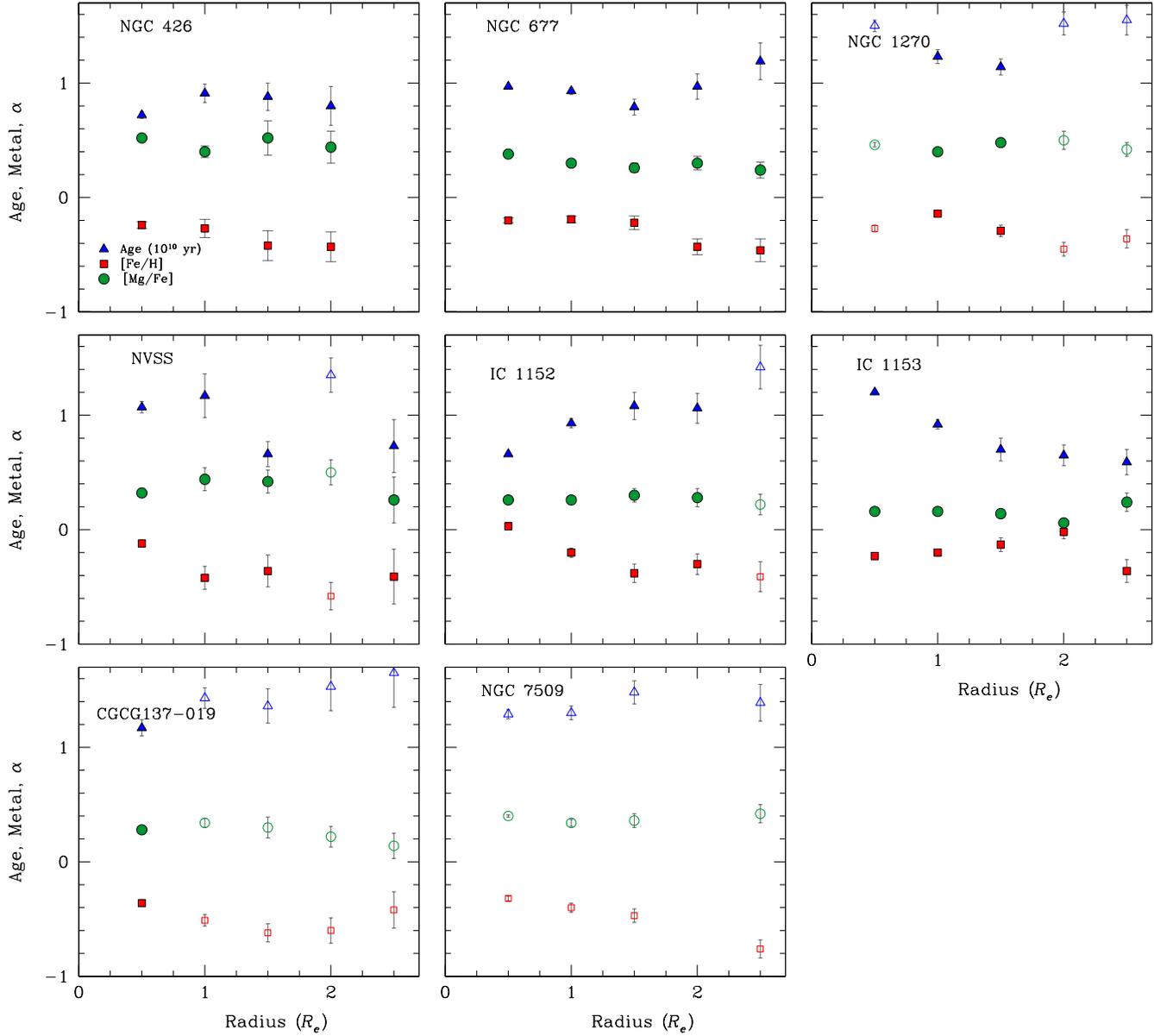,width=0.95\textwidth,keepaspectratio=true,angle=0}
}
\vskip -0mm
\figcaption[]{
  Radial profiles of Age ($10^{10}$ yr; blue triangles), [Fe/H] (red
  squares), and [Mg/Fe] (green circles) for each galaxy.  Radius is measured in
  units of $R_e$.  Open symbols are the adjusted values using H$\beta$+$0.2 \,$\AA, 
  and the two gaps are cases that still fell off of the grid. As above, error bars are
  derived from Monte Carlo simulations. 
\label{fig:radialagefe}}
\end{figure*}

We also use the code {\it EZ\_Ages} by 
\citet[][see also Schiavon 2007]{gravesschiavon2008}.  Here, the age,
metallicity, and alpha abundance ratios are fit iteratively using the full
suite of Lick indices.  {\it EZ\_Ages} solves for the best-fit
parameters by taking pairs of measured quantities (e.g., \hbeta\ EW
and \feave) and then locating the measurements in a grid of model
values spanning the full range of age and (in this case) [Fe/H]
abundance of the models.  The model has a hierarchy of measurement
pairs that it considers, first pinning down age and [Fe/H], then
looking at [Mg/Fe] and so on.  The code then iterates to improve the
best fit values.  Many more elemental abundances can be fitted by {\it
  EZ\_Ages}, thus enabling study of the independent variability of
[N/Fe], [C/Fe], etc.  There is evidence that individual \alp\ elemental ratios
vary independently in individual Milky Way stars
\citep[e.g.,][]{fulbrightetal2007} and possibly in galaxies as well
\citep[e.g.,][]{kelsonetal2006,schiavon2007}.  We do not have adequate
S/N in the blue indices to derive other elemental 
abundances \citep[e.g.,][]{yan2011}, so we just assume
that [Mg/Fe] tracks [$\alpha$/Fe]. 
\vbox{ 
\vskip +8mm
\hskip -3mm
\psfig{file=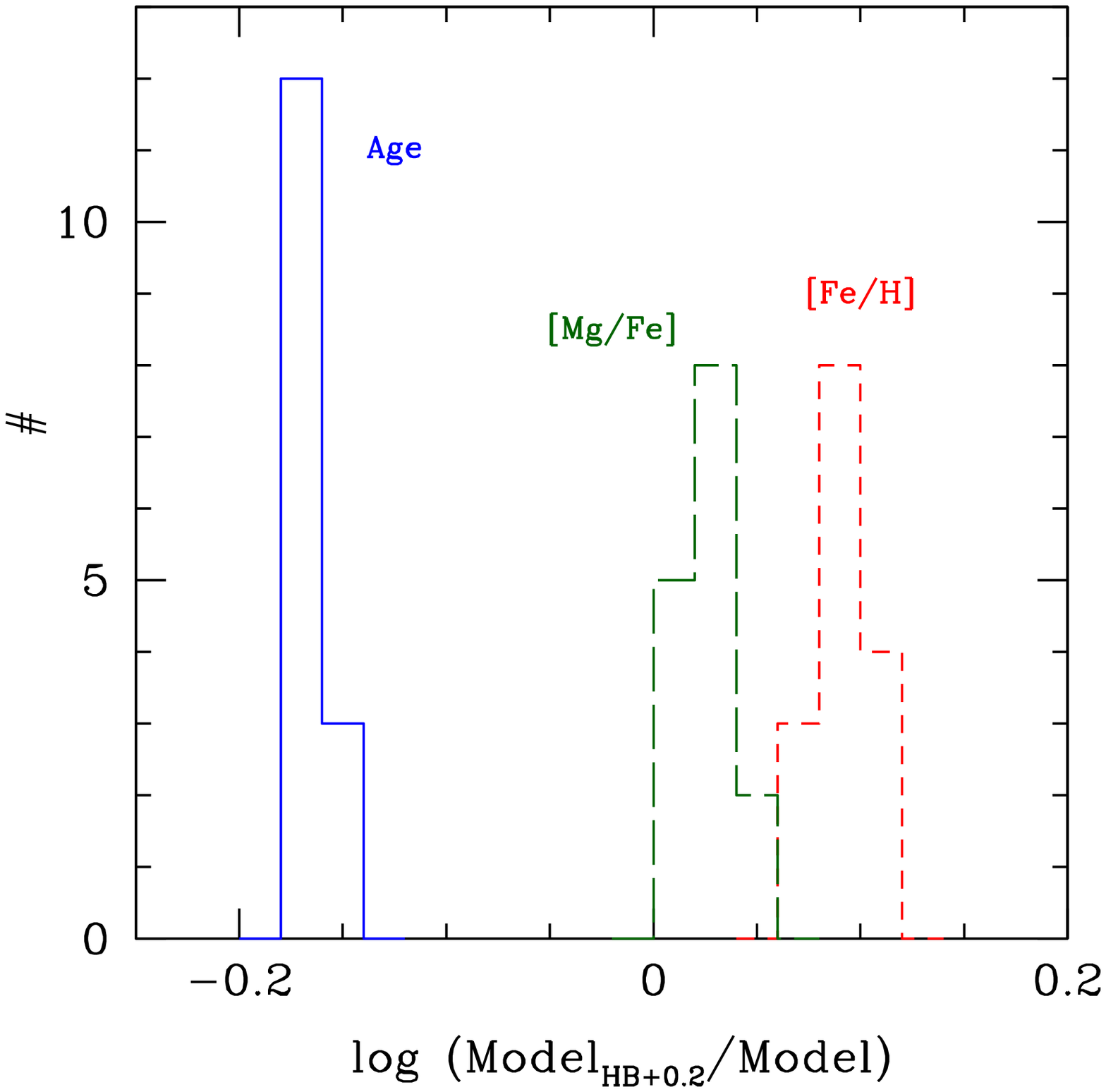,width=0.45\textwidth,keepaspectratio=true,angle=0}
}
\vskip -0mm
\figcaption[]{
  The difference between model parameters from {\it EZ\_Ages} from the
  observed data and when the H$\beta$ index is boosted by $0.2 \,$\AA.  When 
  H$\beta$ is artificially increased, the ages (blue
  solid histogram) are lower by $31 \pm 5$\%, [Fe/H] (red dashed
  histogram) increases by $0.09 \pm 0.01$ dex and [Mg/Fe] (green
  long-dashed histogram) increases by $0.02 \pm 0.01$.  When the 
  \hbeta\ EW is too low and falls off of the grid, these corrections 
  are used to put the model parameters based on \hbeta+$0.2 \,$\AA\ on the 
  same scale.
\label{fig:hbhist}}
\vskip 5mm
\noindent
For the remainder of the paper we will use 
[\alp/Fe] to refer to the \alp-abundance ratios collectively, bearing in mind 
that we have directly measured [Mg/Fe].

The Thomas et al.\ (TMB) approach and the Graves \& Schiavon (GS) model
have somewhat different philosophies, but are inherently similar. 
Both are based on the inversion of single-burst model grids. Both
adjust their primary models for variable $\alpha$-abundance ratios at
a range of metallicities.  TMB use solar isochrones
\citep{cassisietal1997,bonoetal1997} but then modify the indices using
the response functions of \citet{tripiccobell1995}.  GS use solar
isochrones from \citet{girardietal2000} and \alp-enhanced isochrones
from \citet{salasnichetal2000}, with the response functions of
\citet{kornetal2005}.  TMB invert a small number of high S/N indices,
while GS rely on all measured indices in an iterative fashion.  In
principle we can test some of the systematics of the modeling by
comparing our results from the two different approaches.

Finally, note that \citet{gravesschiavon2008} parametrize their models
in terms of [Fe/H] rather than total metallicity [Z/H], so that they
report direct observables.  As described in \citet[][and references
therein]{schiavon2007}, we are not able to measure [O/H] directly, and
thus cannot truly constrain [$Z$/H].  To compare the two models, we
will use the standard conversion
\citep{tantaloetal1998,thomasetal2003}:

\begin{equation}
{\rm [Fe/H] = [Z/H] - 0.94 [\alp/Fe]}
\end{equation}

In Figure \ref{fig:cftmbgraves} we compare [Fe/H] from Thomas et
al. ([Fe/H]$_{\rm TMB}$) with that from Graves \& Schiavon
([Fe/H]$_{\rm GS}$) and likewise for [Mg/Fe], using observations in
all radial bins for each galaxy. The agreement is reasonable.
Overall, we see a scatter of $\langle$[Fe/H]$_{\rm GS}$-[Fe/H]$_{\rm
  TMB} \rangle = -0.06 \pm 0.1$ for [Fe/H] and $\langle$[\alp/Fe]$_{\rm
  GS}$-[\alp/Fe]$_{\rm TMB} \rangle = -0.02 \pm 0.08$ for
  [$\alpha$/Fe].  For the rest of the paper we will focus on the {\it
  EZ\_Ages} results from Graves \& 
Schiavon, but trust that our results can be directly compared with
many in the literature.  Also, we have rerun the {\it
  EZ\_Ages} modeling with the solar isochrones, and the derived
metallicities and \alp-abundance ratios agree within the measurement
errors. The radial dependence of the derived quantities for each galaxy is shown in
Figure \ref{fig:radialagefe}.

\subsubsection{Low \hbeta\ Equivalent Widths}

As mentioned above, there is very likely real but undetectable levels
of \hbeta\ emission that slightly lowers the observed \hbeta\ 
EWs\footnote{We have also investigated whether a non-Gaussian 
line-broadening function or changing spectral resolution could lead 
to \hbeta\ infill.  We broaden the Bruzual \& Charlot models with the 
appropriate line-broadening function, and find that the measured indices only change 
at the $\sim 0.05 \,$\AA\ level.  As described above, in principle sky 
subtraction could cause errors at the 0.1~\AA\ level, but it is hard to 
understand how those errors would be so systematic.  We conclude that 
low-level emission is the most likely culprit.}.  At
the levels measured by \citet{gravesfaber2010} in composite SDSS
spectra ($\sim 0.2 \,$\AA), the age errors are $\sim 2$ Gyr in general
\citep{schiavon2007}.  In some cases, we cannot derive reasonable
model parameters because the measured \hbeta\ EW is too low to fall
onto the SSP grids.  Since we do not have the S/N needed to correct
our spectra on a case by case basis, we use the following procedure to
derive model parameters at the radial bins where the \hbeta\ index
fall off the bottom of the grid.  Note that these corrections are not 
strictly correct, since the level of emission must vary with radial distance.
However, they are the best that we can do at present.

We recalculate the age, \feh, and \mgfe\ for all galaxies with
the \hbeta\ EW increased by $0.2 \,$\AA.  We then derive an average difference
in each measured property between the two sets of models, as shown in
Figure \ref{fig:hbhist}.  The differences in \mgfe\ and \feh\ are very
small and (crucially) show no trend with radius, S/N, or \hbeta\
index. The run with increased \hbeta\ EW returns ages that are $31 \pm
5\%$ dex lower, [Fe/H] values that are $0.09 \pm 0.01$ dex higher and
[\alp/Fe] values that are $0.02\pm 0.01$ dex higher on average than
the unadjusted data.

We correct the model parameters derived from the \hbeta+$0.2 \,$\AA\ run to
align with the fiducial models using the corrections listed above.  At
each radial bin where we could not derive model parameters using the
fiducial \hbeta\ EWs, we instead utilize the corrected model
parameters from the \hbeta+$0.2 \,$\AA\ run.  In the following, all such
points are indicated with open, rather than filled, symbols.  With
this correction, we derive SSP properties for all but two radials bins
over all of the galaxies.  Again, applying the same \hbeta\ correction
for all points that fall of the grid is not strictly correct, since there is 
likely radial dependence in the amount of infill.  
Therefore, we again emphasize that our main strength is in
measuring the radial trends rather than the absolute values of the
stellar population parameters.  Note that \citet{kelsonetal2006} take
a similar, although perhaps more nuanced, approach by shifting all of
their model grids to match the Lick indices of their oldest galaxies.

\section{The Ages and Metal Contents of Stellar Halos}
\label{sec:agemetal}

The most striking trend in Figure \ref{fig:spec_pg1} is the clear and
steady decline in the \mgb\ index out to large radii.  While the
gradients vary from object to object, the qualitative behavior is the
same for all systems.  In contrast, both the $\langle$Fe$\rangle$
index and the \hbeta\ index are generally consistent with remaining flat over
the entire radial range.  In Table 3 we show the gradients in \hbeta,
$\langle$Fe$\rangle$, and \mgb\ measured as $\delta$X $\equiv \delta$
log X$/\delta$ log $R/R_e$ for each index ``X''.  

\vbox{ 
\vskip +8mm
\hskip -4mm
\psfig{file=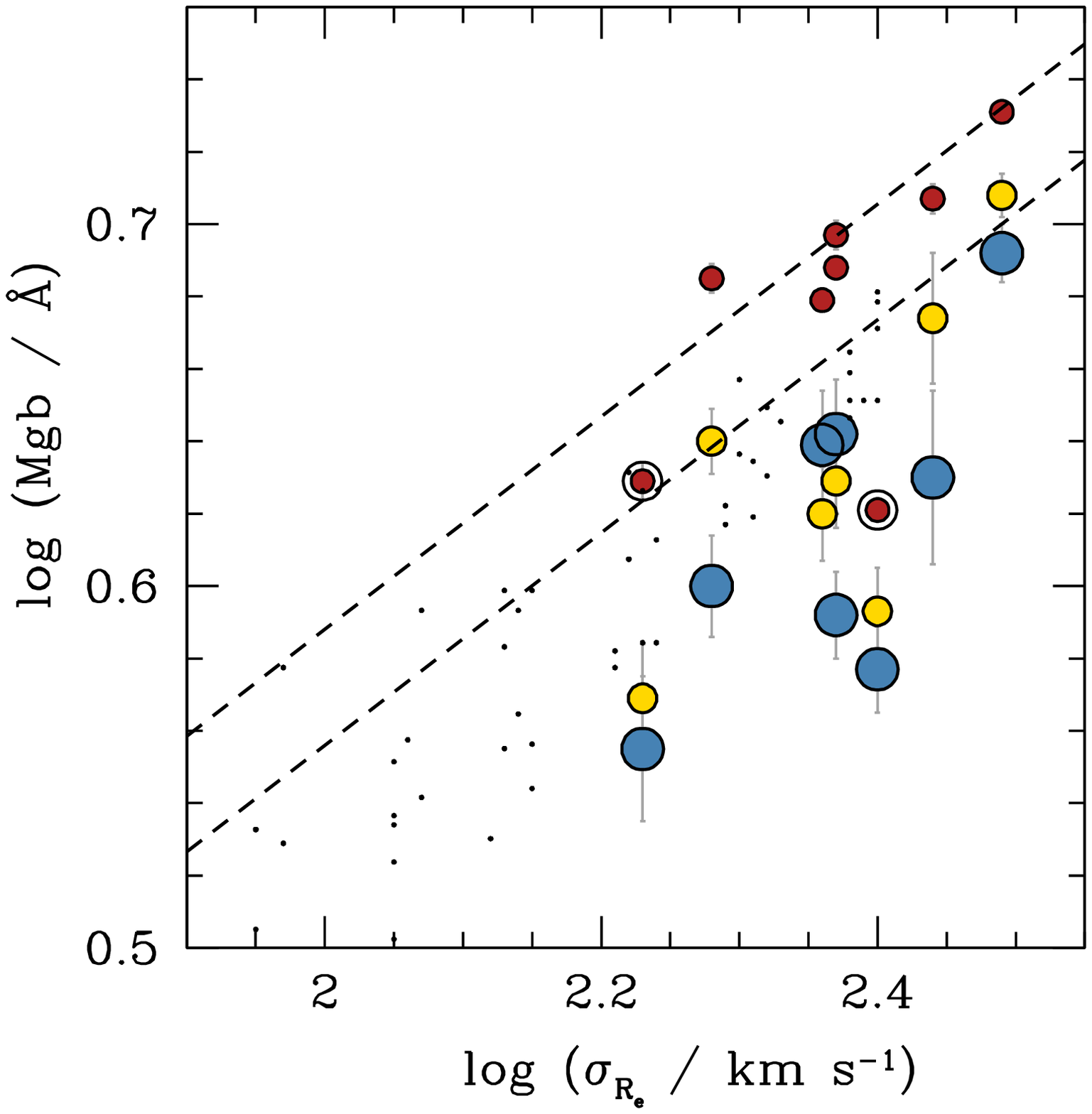,width=0.47\textwidth,keepaspectratio=true,angle=0}
}
\vskip -0mm
\figcaption[]{
  Correlation between the \mgb\ index and central stellar velocity dispersion.
  For direct comparison with the literature, we use the stellar
  velocity dispersion measured within an effective radius, although
  the differences in dispersion are small as a function of radius (see \S 5.1 below).
  We plot measurements for each galaxy at three radii: the central
  0.5$R_e$ (small red filled 
  circles), between 1.5 and 2 $R_e$ (medium yellow circles), and
  between 2 and 2.5 $R_e$ (large blue circles).  The S0 galaxies CGCG137-019 
  and IC 1153 are indicated with a double circle on the central (red) point.  
  For reference, we
  show the measurements from \citet{gravesetal2009} as small dots.
  Since these are SDSS galaxies at $z\approx 0.1$, the light comes
  from $\sim R_e$.  We also show the average relation
  from \citet{trageretal2000} as dashed lines, which show the trend line well within
  the effective radii of average elliptical galaxies.  In terms of \mgb\ EW, the halos of these
  elliptical galaxies have a similar chemical makeup as galaxies of
  much lower mass. 
\label{fig:mgbsig}}
\vskip 5mm

We now ask whether the well-known \mgb-\sigmastar\ relation
\citep[e.g.,][]{benderetal1993}, is preserved at large radii.  In
Figure \ref{fig:mgbsig} we plot the \mgb\ index at $0.5 R_e$, $1.5
R_e$, and $2.5 R_e$ as a function of galaxy stellar velocity
dispersion measured within the effective radius.  This figure visually
displays two interesting trends. First of all, the \mgb\ EWs beyond $2
R_e$ in these massive elliptical galaxies fall significantly below the
central \mgb-\sigmastar\ relation.  Matching the \mgb\ EWs at $2 R_e$
with the centers of smaller elliptical galaxies suggests that the
halo stars were formed in smaller systems.  In \S 5.2 we find that
if these stars were accreted from smaller elliptical galaxies, they
would come in $\sim 10:1$ mergers.  Of course, the more detailed
abundance patterns of the halo stars will give us more clues as to the
possible origins of these halo stars.

Secondly, we do see hints of a \mgb-\sigmastar\ correlation even at
large radii, but with much more scatter.  There is also an intriguing
hint that the slope of the \mgb-\sigmastar\ relation changes. Unfortunately, the
correlation is driven to a large degree by the galaxy NGC 1270, which
has the largest \sigmastar\ value in our sample.  NGC 1270 also
happens to be our only cluster galaxy.  Since the distribution of merger mass ratios 
is nearly independent of mass \citep{fakhourietal2010}, we might expect 
the mass of the typical accreted system to rise with mass, thus
preserving an \mgb-\sigmastar\ relation at large radius. However, we will need a
larger sample at the highest velocity dispersions to say for certain
whether a \mgb-\sigmastar\ trend continues in the galaxy outskirts.

The next obvious question is whether the \mgb\ EW drops primarily
because of changes in \feh\ or [\alp/Fe].  As outlined in the
introduction, all evidence suggests that metallicity decline is the
primary cause for the decline in \mgb\ EW.  Since very little data
exists at such large radii, however, it is worth investigating the
\feh\ and [\alp/Fe] measurements directly.  Again we emphasize that
the absolute values of the derived \feh\ and [\alp/Fe] are unreliable,
but that the gradients should be robust.  Below we present gradients
in the derived metallicities and abundance ratios in order to
determine what drives the striking decline in \mgb\ EW.

\subsection{Gradients}

\vbox{ 
\vskip 5mm
\hskip 0.in
\psfig{file=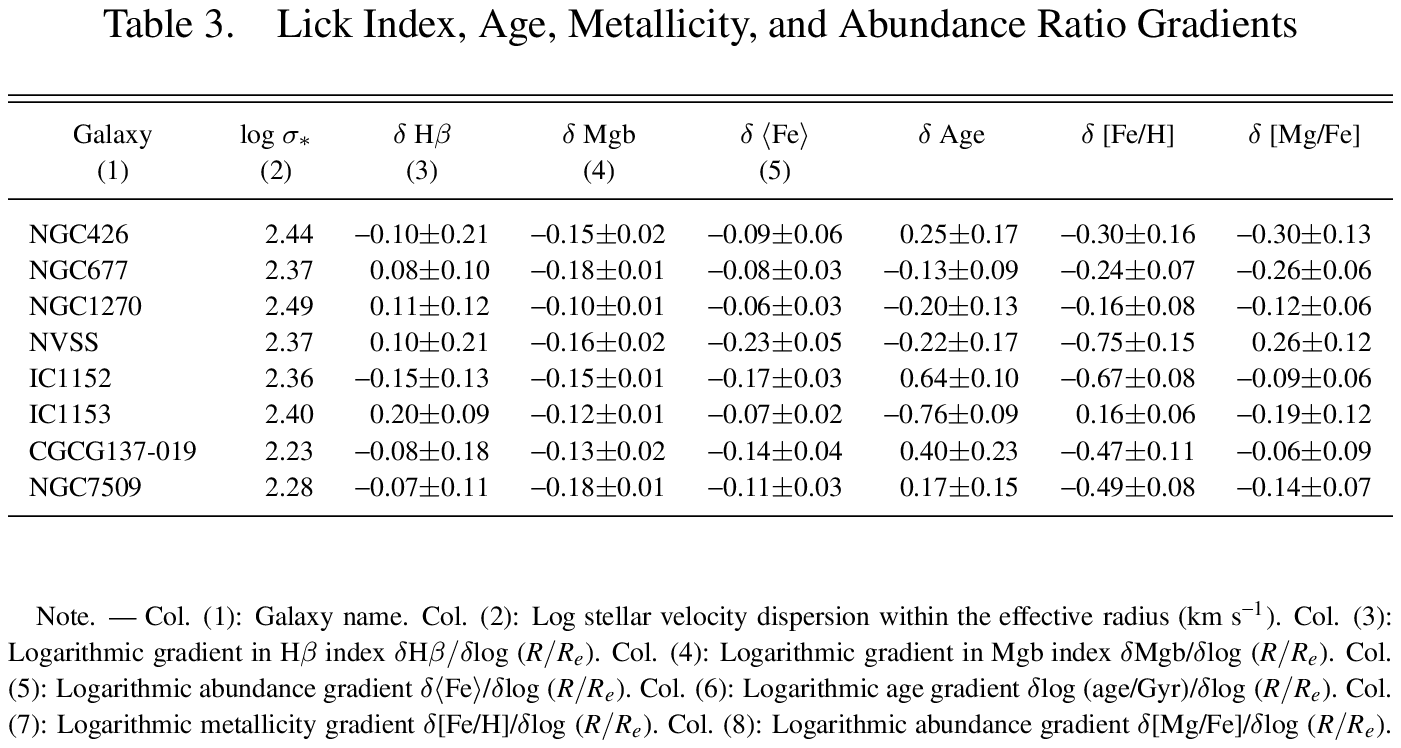,width=0.45\textwidth,keepaspectratio=true,angle=0}
}
\vskip 4mm

As described above, we measure the gradients $\delta$Age, $\delta$\feh, and
$\delta$\mgfe\ as $\delta$X $\equiv \delta$ log X$/\delta$ log
$R/R_e$.  We use adjusted values of age, \feh, and \mgfe\ for objects
that fell off of the grid due to low \hbeta\ EW.  We do a very simple
least-squares fit to the measured quantities as a function of
logarithmic radii in units of $R_e$ (Table 3).  Only \feh\ shows
significant evidence for a significant radial gradient. 
On average, the [\alp/Fe] ratio declines very gently with radius, but the 
trend is not significant in individual cases. 
The age gradients take both positive and negative values, but are rarely 
significant.  Given the uncertainties with \hbeta\ described above, we will 
focus exclusively on metallicity and abundance ratio gradients here.

We should note that unlike most observations in the literature, our
fits are weighted towards the outer parts of the galaxies, and we have
less spatial resolution in the galaxy centers.  We are thus less
susceptible to stellar-population variations in the central regions
caused by late-time accretion and/or star formation.  Indeed,
\citet{baesetal2007} find a clear break in the slopes of metallicity
gradients at small radii, with the gradients getting shallower at
larger radii, as do \citet{coccatoetal2010} for the brightest cluster
galaxy in Coma, NGC 4889.  Our mild metallicity gradients are similar
to those seen at large radius by these authors, as well as by
G. Graves \& J. Murphy in preparation in M~87.

\begin{figure*}
\vbox{ 
\vskip 20mm
\hskip 5mm
\psfig{file=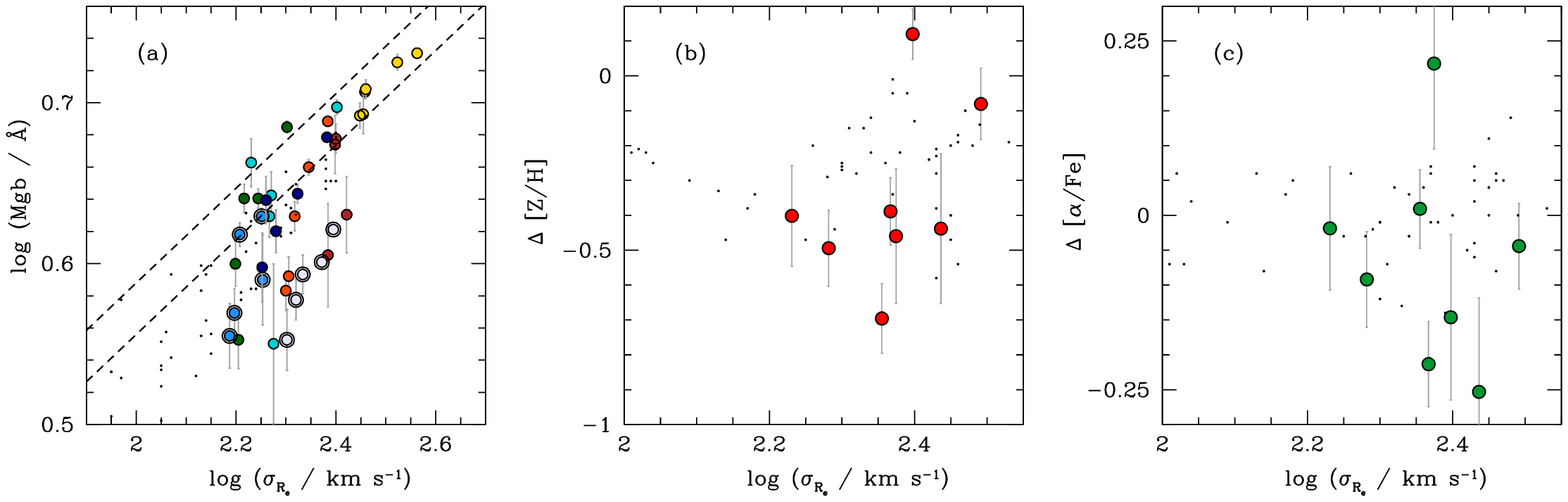,width=0.3\textwidth,keepaspectratio=true,angle=-90}
}
\vskip -0mm
\figcaption[]{
  ({\bf a}): The \mgb\ index as a function of the stellar velocity
  dispersion measured at each radial annulus rather than within $R_e$.
  Each galaxy is shown as a different colored circle. The correlation
  between \sigmastar\ and \mgb\ in individual galaxies as a function of
  radius is considerably steeper than that seen across galaxy centers.
  As in Figure \ref{fig:mgbsig} above, we show the measurements from
  \citet{gravesetal2009} as small dots.  Since these are SDSS galaxies
  at $z\approx 0.1$, the light comes from $\sim R_e$.  We also show
  the average relation from \citet{trageretal2000}, which shows the
  trend line well within the effective radii of average elliptical
  galaxies.
({\bf b}): Gradients in metallicity $\delta$[Z/H]/$\delta$ log ($R/R_e$) 
as compared with the stellar velocity dispersion of the galaxy.  Our galaxies are 
shown as large red circles.  We compare with 
the compilation of \citet{spolaoretal2010}, and find reasonable agreement in the 
velocity dispersion range covered by our data.
We measure \sigmastar\ within $R_e$, while Spolaor et al. use $1/8 R_e$, but the 
gradients in \sigmastar\ over these radii are generally small 
\citep[e.g.,][]{jorgensenetal1995,cappellarietal2006}. Also note that 
Spolaor et al.\ tabulate gradients out to the effective 
radii in most cases, while we measure them between $0.5 R_e$ and $2.5 R_e$.  
({\bf c}): Same as (b) for gradients in $\delta$[$\alpha$/Fe]/$\delta$ log ($R/R_e$) 
(big green circles).  Again note that [Mg/Fe] is assumed to trace [$\alpha$/Fe].
\label{fig:cfspolaor}}
\end{figure*}

In principle, correlations between the stellar population gradients
and other properties of the galaxy provide additional clues as to the
origin of the gradients.  We examine the relation between \sigmastar\
and gradients in Figure \ref{fig:cfspolaor}.  Our results are
consistent with the \citet{spolaoretal2010} result that the gradients
in elliptical galaxies show a larger scatter at higher \sigmastar.  In
our small sample, we do not find a correlation between the decline in
metallicity and the isophote shape, with the metallicity gradients
being substantially shallower than the decline in isophote level with
radius.  We do not find a correlation between gradients in stellar
velocity dispersion and gradients in indices or gradients in
metallicity or abundance patterns, but the sample is yet small.
Eventually it will be interesting to look for correlations between the
local escape velocity and the metallicity gradient as may be seen if
the metallicity is set locally by the ability of gas to escape the
galaxy \citep{scottetal2009,weijmansetal2009}.

\vskip 5mm

\subsection{Where do the Halo Stars Come From?}

Figure \ref{fig:mgbsig} presents the intriguing possibility that we
may uncover the mass scale of accreted satellites by matching the
metallicities and abundance ratios of the stellar halos with the
centers of smaller elliptical galaxies. In this section we take the
measured gradients in \mgb\ index, \feh, and \mgfe\ and attempt to
constrain the typical mass of an accreted satellite.  As we will show,
the abundance patterns in the halo stars do not match the central
regions of any local elliptical galaxies, thereby complicating our
efforts to find the progenitors of the halo stars.  Nevertheless, we
can derive an approximate mass ratio from our measured gradients,
accepting that present-day galaxies do not form perfect analogs of the
accreted satellites.

Using our observed \sigmastar\ within $R_e$, we assign a central value 
of abundance ratio and metallicity using the
\mgb-\sigmastar, \feh-\sigmastar, and \mgfe-\sigmastar\ relations
from \citet{gravesetal2007}.  We then use our observed gradients to
link the stars at large radii with the \sigmastar\ of its most likely
progenitor.  For instance, we assign an \feh\ to each galaxy using the
\feh-\sigmastar\ relation from \citet{gravesetal2007}. Then we use our
measured gradients to calculate \feh\ at $2-2.5 R_e$.  That \feh\
value at large radius is matched to the \sigmastar\ of an accreted
galaxy with the same metallicity using the \feh-\sigmastar\
relation. The derived value of \sigmastar\ is translated into a
stellar mass $M^*$ using the projections of the Fundamental Plane
presented by \citet{desrochesetal2007}.  The typical mass of accreted
galaxies based on each property is shown in Figure
\ref{fig:calsiggal}.

Figure \ref{fig:calsiggal} quantifies the typical mass of accreted
satellites.  The figure strengthens our conclusion from Figure
\ref{fig:mgbsig}.  Based on the \mgb\ EWs alone, we find that the
stars at $>2R_e$ were accreted from galaxies $\sim 10$ times less
massive than our target galaxies (blue points).  However, the figure
makes very clear that the metallicities and abundance ratios of our
stellar halos cannot be matched with present day elliptical galaxies
of any mass.  Looked at another way, it says that the \feh\ gradients
we see in individual halos are steeper than the \feh-\sigmastar\
relation for the population overall, while the \mgfe\ gradients are
shallower (Figure \ref{fig:cfspolaor}).  The \mgb-\sigmastar\ relation
in the galaxy outskirts taken alone suggests that stellar halos are
built by $\sim 10:1$ mergers, but the halo stars have lower
metallicities and higher \alp-abundance ratios than present-day
low-mass ellipticals.  

Of course, accreted satellite systems need not be small elliptical
galaxies, and may well have started their lives as gas-rich disk
galaxies.  Thus, it is useful to consider spiral galaxies as well.
Since disks also have low surface brightness, there are few studies of
abundance patterns in disks based on stellar absorption features
\citep[][]{gandaetal2007,yoachimdalcanton2008}.  Furthermore, the
interpretation using SSP models is severely complicated by the clear
ongoing star formation in these systems \citep[][]{macarthuretal2009}.
There is general agreement that massive red bulges show similar
patterns and scaling relations as elliptical galaxies
\citep[e.g.,][]{moorthyholtzman2006,robainaetal2012}.  At lower mass,
\citet{gandaetal2007} find that even the bulge regions of later-type
spirals have younger ages, lower metallicities and solar abundance
ratios compared to elliptical galaxies.

Due to their protracted star formation histories, present-day
late-type spirals do not share the high abundance ratios of the
stellar halos studied here.  Instead, it may be more productive to
consider individual components of present-day galaxies.  For instance,
thick disks are clearly older than thin disks, and depending on their
formation channel may share characteristics of these stellar halos.
However, aside from our own galaxy, it remains very difficult to
obtain robust stellar abundance patterns in thick disks
\citep{yoachimdalcanton2008}.  In Figure \ref{fig:cfmw} we show the
abundance ratios at $2-2.5 R_e$ in our sample as derived using the
central relations from Graves et al. and our gradients.  We compare with other 
stellar populations including elliptical and spiral galaxies, and subcomponents 
of our own galaxy. This figure
shows that some of our galaxy halos have similar abundance patterns as
the Milky Way thick disk while others are more consistent with
low-mass elliptical galaxies. We now address plausible scenarios for
how the stellar halos were assembled.

\section{Discussion}

\subsection{Theoretical Expectation}

We have established clear gradients in \mgb\ EW with radius in all of
the galaxies in our study.  Based on Lick index inversion methods we
have argued that in general these gradients are dominated by
metallicity with a weak contribution from abundance ratio gradients as
well.  While the \mgb\ EWs of the galaxy halos match those of galaxies
an order of magnitude less massive, the stars appear to have lower
metallicity and higher \alp-abundance ratios than do low-mass ellipticals.
We now review the various physical processes that we believe can
impact the observed chemistry, in order to determine which scenarios are
favored by our observations. 

We start with ``monolithic'' collapse, by which some large fraction of
the galaxy is built in a single dissipational burst of star formation
\citep[e.g.,][]{eggenetal1962}.  While we believe that we live in a
hierarchical Universe in which large galaxies are built up through the
merging of smaller parts, the central stellar populations of massive
ellipticals clearly imply that their stars were formed rapidly at
redshifts $z > 2$ \citep[e.g.,][and references
therein]{thomasetal2005}.  A rapid dissipational phase at high
redshift is likely an important part of elliptical galaxy formation,
followed by late-time dry merging
\citep[e.g.,][]{taletal2009,vandokkumetal2010,newmanetal2011}.  There
is strong theoretical support for such a ``two-phase'' picture
\citep[e.g.,][]{naabetal2009,oseretal2010} and high-redshift
progenitors, in which star formation has ceased at early times, are
observed \citep[e.g.,][]{krieketal2009}.

A large body of work, starting with \citet{larson1974}, has considered
the chemical evolution of monolithic collapse models, including a
heuristic star-formation law, chemical enrichment, and (typically)
galaxy-scale mass loss driven by supernovae.  Since these galaxies
have deep potential wells, the gas in the center cannot easily be
ejected, but instead is enriched by preceding generations of star
formation and grows metal-rich.  In contrast, in the outer parts of
the galaxy, winds can be effective at ejecting metals.  Steep
gradients in metallicity and \alp\ abundance ensue
\citep[e.g.,][]{carlberg1984,arimotoyoshii1987,kawatagibson2003,kobayashi2004}.

In a modern cosmological context, massive halos that host the
progenitors of massive elliptical galaxies are constantly bombarded
with smaller halos
\citep[e.g.,][]{boylan-kolchinetal2009}. Furthermore, we see galaxies
merging \citep{toomretoomre1972,schweizer1982}.  Mergers will scramble
the orbits of the constituent parts to some degree and wash out
metallicity gradients \citep[e.g.,][]{white1980}.  The degree of
mixing depends on a variety of factors.  In a violent relaxation
scenario (without dissipation) existing stars may not migrate much,
thus preserving the original chemical patterns \citep{vanalbada1982}.
In contrast, gas-rich major merging should efficiently supply gas to
the center of the remnant, where it will form metal-enriched stars and
steepen metallicity gradients
\citep[e.g.,][]{mihoshernquist1996,coxetal2006}.  Modern simulations
that account for cosmological merging and chemical evolution conclude
that merger remnants will have shallower metallicity gradients on
average, with a much larger scatter than the monolithic collapse case
\citep[e.g.,][]{kobayashi2004}.

\subsection{Constraints From Data}

Let us briefly review what the observations tell us (see also the introduction 
for more complete references):

\begin{enumerate}

\item
  There is a strong correlation between \mgb\ EW and \sigmastar\ 
  measured within the effective radius of the galaxy
  \citep[e.g.,][]{faber1973,dressleretal1987,benderetal1993}.  The
  majority of this trend is attributed to metallicity, but there is
  also a trend between \alp-abundance ratio and \sigmastar\
  \citep[e.g.,][]{wortheyetal1992,gravesetal2009}.  Much like the
  mass-metallicity trend observed in star forming galaxies, these
  trends must arise at some level because of the relative ease of
  ejecting metals from the shallow potentials of low-mass galaxies
  \citep[e.g.,][]{dekelsilk1986,tremontietal2004}.

\item 
  In $\sim L*$ elliptical galaxies, the metallicity decreases
  outwards gently, falling by $\sim 0.1-0.3$ dex per decade in radius.
  The gradients are too shallow in general to agree with pure
  monolithic collapse scenarios. No clear trends are seen in
  \alp-abundance ratio gradients, with increasing, decreasing and flat
  trends observed \citep[e.g.,][]{kuntschneretal2010}, again in
  disagreement with monolithic collapse. In more massive elliptical
  galaxies, metallicity gradients still dominate, but there is a wider
  dispersion in the gradient slopes at a given \sigmastar\
  \citep{carolloetal1993,spolaoretal2010}, consistent with what we observe
  (Figure \ref{fig:cfspolaor}).

\item
  We add an additional robust spectroscopic point beyond $2 R_e$.  The
  \mgb\ EW in the stellar halos match the values seen in local
  elliptical galaxies that are ten times less massive than our sample
  galaxies. However, the centers of present-day elliptical galaxies in
  this mass range have more metals and lower values of [\alp/Fe] than
  do the stellar halos.  Our gradient observations are strongly in
  contrast with predictions from monolithic collapse scenarios, in
  which \alp-abundances would decrease outwards
  \citep[][]{kobayashi2004}.  The question is whether we can explain
  the observed stellar properties if the halos were built via minor
  merging at late times.
\end{enumerate}

At first glance, our observed \alp-abundance ratios are difficult to
understand in any scenario.  We rule out a pure monolithic-collapse
scenario because of the lack of gradient in abundance ratios.  Major
mergers are not strictly excluded, but seem unable to produce such
consistent decreasing metallicity gradients without some tuning. If,
in contrast, the outskirts are built up via minor merging at $z < 1$
\citep[e.g.,][]{naabetal2009}, then we would expect the stars to have
the same metallicities and abundance patterns as small elliptical
galaxies today.  Figure \ref{fig:calsiggal} demonstrates that [\alp/Fe]
is too high at a given [Fe/H] to derive from present-day low-mass 
elliptical galaxies.

\vbox{ 
\vskip 8mm
\hskip -5mm
\psfig{file=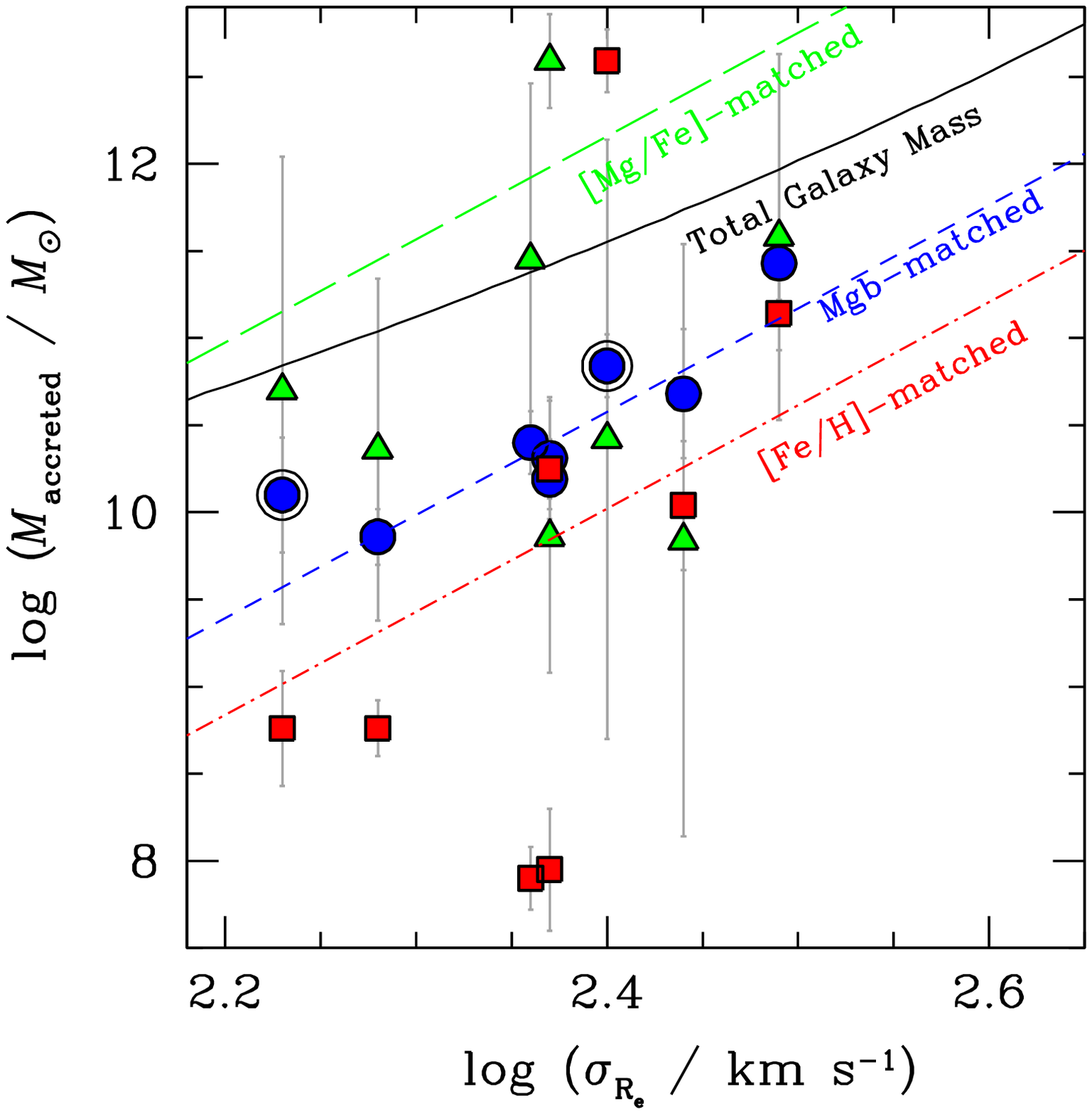,width=0.47\textwidth,keepaspectratio=true,angle=0}
}
\vskip -0mm
\figcaption[]{
  The characteristic mass of an accreted satellite at $2.5 R_e$ as inferred from 
  the \mgb\ EW (blue circles), \feh\ (red squares), and \mgfe\ (green triangles). 
  As in Figure \ref{fig:mgbsig}, S0s are indicated with double circles.
  To derive the accreted mass, we assume previously derived \mgb-\sigmastar, 
  \feh-\sigmastar, and \mgfe-\sigmastar\ relations \citep{gravesetal2007} 
  to assign central values of these quantities to our galaxies. Using our 
  observed gradients, we match the abundance patterns beyond $2 R_e$ 
  to a present-day elliptical galaxy with the same \mgb\ EW, \feh, or \mgfe.
  We translate the \sigmastar\ value to a stellar mass using the Fundamental 
  Plane relations from \citet{desrochesetal2007}.  The total galaxy mass derived 
  using the Fundamental Plane is shown with the solid black line.  
  We fit average relations between \sigmastar\ and $M_{\rm accreted}$ based 
  on each measurement with a fixed slope to guide the eye only
  (\mgb\, dashed blue line; [Fe/H], dot-dashed red line; 
  [Mg/Fe], long-dashed green line).
  Our indirect technique allows us to circumvent uncertainties in the absolute 
  values of \feh\ and \mgfe.  If the stellar halos were constructed 
  from analogs of present-day ellipticals, then the accreted mass derived from 
  each indicator would agree.  Instead we find that present-day ellipticals cannot 
  simultaneously match the observed low values of \feh\ and high values of \mgfe.
\label{fig:calsiggal}}
\vskip 5mm

It is useful to draw an analogy with studies of the Milky Way
halo.  Early suggestions that the halo may be built by the accretion
of satellites \citep{searlezinn1978} were called into question by the
observation that the abundance patterns of stars in the Milky Way halo
do not match those of the existing satellites \citep[e.g.,][ and
references therein]{vennetal2004}.  However, satellites that are
accreted early by the Milky Way will have a truncated star-formation
history.  Thus, they will have high [\alp/Fe] ratios compared to
systems that continue to accrete gas and form stars until the present
day \citep[e.g.,][]{robertsonetal2005,fontetal2006}.
\citet{tisseraetal2011} track the chemical evolution of eight
Milky-Way analogs with hydrodynamical simulations.  Most of the mass
in each simulated halo is accreted (much of it after a redshift of
one) rather than formed in situ.  The accreted stars are metal poor
and \alp-enhanced, since the stars were predominantly formed at early
times in small halos that have truncated star formation histories.  We
observe the same trend in the halos of the massive early-type galaxies
examined here.

Our galaxies are considerably more massive than the Milky Way, and for
the most part they do not have disks at the present time.  There must
be many differences in their merger histories from that of the Milky
Way, and the absolute metallicities of the Milky Way halo stars are
much lower (Fig. \ref{fig:cfmw}).  Nevertheless, we suggest that the
halos in both cases are built by the accretion of smaller galaxies
before these small systems have a chance to self-enrich.  Thus, we
support a scenario in which massive elliptical galaxies were built up
via gas-free minor merging at late times
\citep[e.g.,][]{naabetal2009}.  The abundance patterns of the massive
halos differ from those seen in $\sim L^*$ ellipticals today because
the former had their star formation history truncated when they were
accreted.  In contrast, we know that $\sim L^*$ ellipticals had some
ongoing star formation at late times \citep[e.g.,][]{babulrees1992,
  thomasetal2005,kolevaetal2011}.

The simulations presented in \citet[][see also C.~Lackner et al.  in
preparation]{oseretal2010} provide strong support for halo build-up
via the late accretion of small satellites.  They consider galaxies
with $M*$ ranging from $5 \times 10^{10}-4 \times 10^{11}$~\msun, and
find that half of the stellar mass was accreted after a redshift of
$z \approx 1$.  They are able to reproduce the observed size evolution
in elliptical galaxies \citep[e.g.,][]{vanderweletal2008}.
Furthermore, they reproduce the observed increase in stellar mass on
the red sequence over the last eight billion years
\citep[e.g.,][]{faberetal2007}.  Of more direct importance to our
story, while accreted late, the majority of the accreted stars were formed at $z \gtrsim
3$.  These stars were by necessity formed rapidly out of
low-metallicity gas.  Of course, our sample is still small, but
eventually we hope to have a large enough sample to look for
differences in elliptical halo abundance patterns as a function of
mass.

At a redshift of $z\approx1$, the progenitors of local $L^*$
ellipticals could well have consisted of old, metal-poor, and
\alp-enhanced stars.  In the schematic picture of
\citet{thomasetal2005}, $\sim L^*$ ellipticals are forming the
majority of their stars over a few Gyr around $z \lesssim 1$.  If
roughly half of the stars are formed at $z < 1$, with a timescale of
$\sim 3$ Gyr, while the original population has [\alp/Fe]=0.2-0.3,
the final mass-weighted abundance would be [\alp/Fe]$\approx$0.15, as
observed for $L^*$ ellipticals today.  Thus, those galaxies that are
not cannibalized by larger galaxies can easily self-enrich to form 
the observed populations today. We make a testable prediction of strong
evolution in the metallicity and [\alp/Fe] ratios of $\sim
L^*$ elliptical galaxies from $z \approx 1$ to the present.

In our analysis, we have assumed a constant initial mass function.  If
instead massive elliptical galaxies have either a top-heavy
\citep{vandokkum2008,dave2008} or a bottom-heavy
\citep{vandokkumconroy2011} initial mass function compared to
lower-mass systems, that may change the interpretation of the
gradients.  We have also ignored the possible variation between 
different \alp\ elements, particularly nitrogen \citep{schiavon2007}.

\section{Summary}

We have used the Mitchell Spectrograph to gather high
S/N spectra of eight massive early-type galaxies out to $2.5 R_e$, in
order to study the chemistry of their stellar halos.  
Looking first at the trends in Lick indices with radius, we
find that the EW of \mgb\ drops, such that the well-known
\mgb-\sigmastar\ relation is not preserved at large radii.  Instead
the \mgb\ EWs at large radii are similar to those found in the centers
of galaxies that are an order of magnitude less massive.

We show that the well-known metallicity gradients seen within $R_e$ 
continue to the furthest radii probed here.  In
contrast, [\alp/Fe] does not drop substantially in any object, and
certainly never approaches the solar value.  Thus, the stars in the
outer regions of these elliptical galaxies are metal-poor and
\alp-enhanced, much like the stars in the Milky Way halo.  We suggest
that the outer parts of these galaxies are built up via minor merging 
with a ratio of $\sim 10:1$,
but that the accreted galaxies did not have sufficient time to lower their
\alp-abundance ratios to those seen in $\sim L^*$ elliptical galaxies
today.  

\vbox{ 
\vskip +8mm
\hskip -4mm
\psfig{file=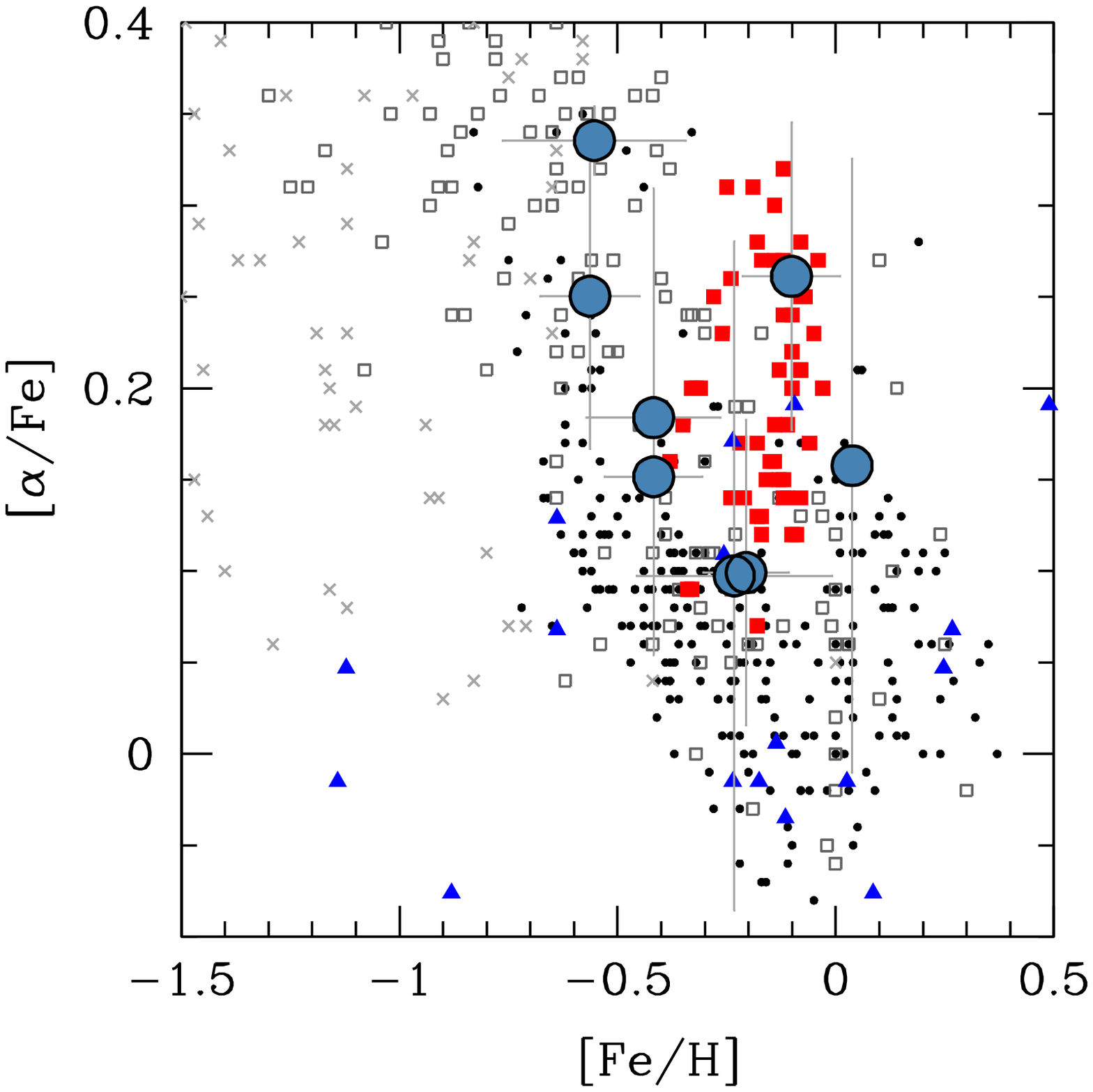,width=0.47\textwidth,keepaspectratio=true,angle=0}
}
\vskip -0mm
\figcaption[]{
We infer the \feh\ and \mgfe\ values at $2-2.5 R_e$ using the central relations of 
\citep{gravesetal2007} and our measured gradients (large blue circles). 
We compare the abundance ratios and metallicites in our stellar halos 
with Milky Way stars from \citet{vennetal2004}, including thin disk 
({\it small black circles}), thick disk ({\it small grey open squares}), and halo 
({\it small grey stars}) stars.  For comparison we also show the track of 
the \citet{gravesetal2009} composite elliptical galaxies from the SDSS 
({\it filled red squares}) and the central regions of late-type spiral bulges 
from \citet[][{\it filled blue triangles}]{gandaetal2007}.  Again, 
taken as a group, our stellar halos are not well-matched by the integrated 
properties of galaxy centers today.  However, we do see some overlap with 
Milky Way thick disk stars and low-mass elliptical galaxies.
\label{fig:cfmw}}
\vskip 5mm
\noindent

This paper is only a proof of concept; the Mitchell Spectrograph is
ideally suited to study the faint outer parts of galaxies, and there
is a considerable amount of follow-up work to be done.  First of all,
we would like to investigate the kinematics in the outer parts of
these galaxies to determine whether there are correlations between
angular momentum content and metallicity.  We are working on gathering
a larger sample, with a full sampling of velocity dispersion, size,
and environment, to see whether the radial gradients in (e.g.,)
metallicity, correlate with the size of the galaxy at fixed
\sigmastar, or the large-scale environmental density.  It seems clear
that galaxy evolution is accelerated in rich environments
\citep[e.g.,][]{thomasetal2005,papovichetal2011}, leaving subtle
imprints in the central stellar populations of galaxies
\citep{zhuetal2010}.  Whether that will leave clear signatures in the
gradients remains unknown.  Even in our own small sample there are
real differences from galaxy to galaxy, and it will be very
interesting to see whether the large-scale environment is the cause.

\acknowledgements

The referee gave us an extremely prompt and thorough report 
that considerably improved this manuscript.
We thank G. Blanc and M. Song for crucial assistance with data
reduction.  We thank G. Graves, J. E. Gunn, L. C. Ho, J. P. Ostriker, and
B. E. Robertson for many stimulating discussions about both the
measurements and the science. J.M.C. is supported by an NSF Astronomy
and Astrophysics Postdoctoral Fellowship under award AST-1102525.

\end{document}